\documentclass[preprint]{elsarticle}

\usepackage{graphicx}
\usepackage{amsmath}
\usepackage[utf8]{inputenc}
\usepackage[lined]{algorithm2e}
\usepackage[english]{babel}

\begin{document}

\begin{frontmatter}

\title{Studying the Impact of Negotiation Environments on Negotiation Teams' Performance}

\author{Víctor Sánchez-Anguix}
\author{Vicente Julián}
\author{Vicente Botti}
\author{Ana García-Fornes}
\address{Universidad Politécnica de Valencia, Departamento de Sistemas Informáticos y Computación, Camí de Vera s/n, 46022, Valencia, Spain, \{sanguix,vinglada,vbotti,agarcia\}@dsic.upv.es}

\begin{abstract}
In this article we study the impact of the negotiation environment on the performance of several intra-team strategies (team dynamics) for agent-based negotiation teams that negotiate with an opponent. An agent-based negotiation team is a group of agents that joins together as a party because they share common interests in the negotiation at hand. It is experimentally shown how negotiation environment conditions like the deadline of both parties, the concession speed of the opponent, similarity among team members, and team size affect performance metrics like the minimum utility of team members, the average utility of team members, and the number of negotiation rounds. Our goal is identifying which intra-team strategies work better in different environmental conditions in order to provide useful knowledge for team members to select appropriate intra-team strategies according to environmental conditions.
\end{abstract}


\end{frontmatter}

\section{Introduction}
\label{sec:introduction}
Agreement technologies \cite{luck08,sierra11} conform an emergent research area among scholars in artificial intelligence and autonomous agent systems. Autonomous software agents act reactively and proactively with the objective of maximizing their human users' goals. Nevertheless, as systems tend to be more complex, so do agents' goals, and agents cannot achieve their goals without the cooperation of other agents. Given the open nature of many multi-agent systems, conflict may be inherent among agents. Hence, distributed mechanisms that allow agents to solve conflict and cooperate are a necessity. Agreement technologies have been actively researched bearing in mind the aforementioned necessity. 

Automated negotiation \cite{kraus97,jennings01,lopes08} is one of the core topics in agreement technologies. Basically, agents in conflict engage in an automatic offer exchange process which gradually leads towards a final solution, or agreement, that solves conflict and makes cooperation among agents possible. The most common use for automated negotiation has been electronic commerce \cite{lomuscio03}, but it should be highlighted that the applicability of this technology has been demonstrated in other domains like collaborative design \cite{klein03}, labor management disputes \cite{sycara93}, and mediation between human negotiation parties \cite{chalamish11}. 

Despite being widely studied by scholars from different disciplines like artificial intelligence, game theory, and social sciences, studies have largely focused on processes whose parties (bilateral, or multiparty) are formed by single individuals \cite{faratin98,ehtamo01,faratin02,serrano03,digiunta06,fatima06,behfar08,halevy08,ito10}. However, some real world scenarios bring about negotiation parties that are formed by more than a single individual. For instance, when an organization negotiates with another organization the selling of a product line, it is usual for organizations to send a group of representatives to negotiate with the other organization. Another example, probably a more quotidian example, involves a married couple that negotiates the purchase of a house with a seller. In this case, the married couple is actually a negotiation party which is formed by two individuals instead than a single individual party. To conclude with the list of real examples, the reader could also think of a group of friends that want to go on a holiday together. This party, conformed by all the friends, has to negotiate a deal with the travel agency if they want to achieve their desired goal. 

This kind of multi-individual party is known in the social sciences as a negotiation team \cite{thompson96,thompson01,behfar08,halevy08}: \textit{a group of interdependent people that join and act together as a single negotiation party because of their shared interests, related to a negotiation}. The rationale behind negotiation teams is mainly twofold. First, team members may have different expertise and negotiation skills that are needed to tackle the negotiation problem successfully. Second, the multi-individual entity that negotiates may be formed by multiple stakeholders with different sub-goals and preferences regarding the final negotiation outcome. We can imagine how an IT company may send a negotiation team formed by experts (different knowledge and skills) from the sales department, marketing department, and R\&D department to successfully negotiate a new project with the local administration, how the wife and the husband may have different opinions with respect to house pricing, location, and facilities, and how each friend may have different interests regarding hotel location, number of days to spend, and pricing regarding their travel. 

Electronic applications, and consequently automated negotiation, are not alien to scenarios that may involve agent-based negotiation teams (ABNT). For instance, group travel e-markets, group buying in e-markets, electronic management of farming cooperatives, negotiation support systems for real human teams, and agent-based simulation may be some of the applications where ABNT may be used. From our point of view, we are interested in ABNT whose members may have different preferences regarding the negotiation issues, and, more specifically, we are interested in models for electronic markets.

In this paper, we present four intra-team strategies for an ABNT that negotiates with a single opponent. Intra-team strategies, also known as team dynamics, govern which decisions are taken as a team, and how and when those decisions are taken \cite{sanchez-anguix10}. The relationship between intra-team strategies and team performance is direct. Hence, it became the focus of our current research. It has been documented that environment conditions such as the deadline, concession speed, and reservation utility may affect the impact of single-individual bilateral strategies \cite{faratin98}. However, in the team case, new conditions like the number of team members, team preferences' diversity, and the emergent effect of aggregating team members' behaviors/actions may also end up affecting team performance. Prior to the negotiation process, negotiation teams face the challenge of selecting which intra-team strategy should be employed. If environmental conditions have an effect on the performance of the different intra-team strategies, the intra-team strategy for the negotiation at hand should be selected accordingly to the current environmental conditions inferred by team members. Our research goal is identifying which intra-team strategies perform better according to different negotiation environments under different team performance measures. The long term goal is employing the results of this article for helping team members to select the proper intra-team strategy. 

Hence, four intra-team strategies that guarantee four minimum levels of unanimity regarding team decisions are presented in this article: representative (no unanimity guaranteed), Similarity Simple Voting (plurality/majority guaranteed), Similarity Borda Voting (semi-unanimity guaranteed), and Full Unanimity Mediated (unanimity guaranteed). Due to the large amount of variables that may affect the negotiation, we employ an empirical approach to study the behavior of the four intra-team strategies. We study and identify which are the most appropriate strategies according to different environmental conditions and team performance measures. This article, is partially based on our previous work regarding intra-team strategies for negotiation teams \cite{sanchez-anguix11,sanchez-anguix12}, where we presented initial results and simulations. In this article, we extend our empirical experiments by incorporating new environmental conditions (i.e., team size, different deadlines), carrying out a more fine-grained analysis of previous environmental conditions (i.e., deadline, concession speeds), and presenting revised versions of the four intra-team strategies.

The article is organized as follows. First, we describe the assumptions of our negotiation model (Section \ref{sec:assumptions}). After that, the details of the four intra-team strategies are thoroughly described in Section \ref{sec:intra}. Then, in Section \ref{sec:exp} the article depicts which negotiation environments and team performance metrics have been studied, and it presents the results and analysis of our experiments. Afterwards, the present work is related to other works in the area of artificial intelligence and automated negotiation (Section \ref{sec:related}). Finally, we briefly state the conclusions of our study and point out some future and interesting lines of work in Section \ref{sec:conclusions}.
\section{General Model Description}
\label{sec:assumptions}
In this section, we describe the assumptions of our model. We have divided the assumptions in two different categories: general assumptions and opponent assumptions. The general assumptions directly affect the nature of the negotiation at hand and are shared between parties (e.g., protocol, number of parties, attribute types, etc.), whereas opponent assumptions describe the strategy carried out by the opponent.
\subsection{Negotiation Setting}
\label{sec:general}
\begin{itemize}
 \item The team $A$ is formed by $M$ different agents $a_{i}, 1\leq i \leq M$. It should be stated that team membership is considered static during the negotiation process. Dynamic ABNT are not considered in this work, and they are appointed as future work.
 \item The common goal of the team $A$ is negotiating a successful deal with the opponent $op$. Thus, in this case we assume an implicit representation of the teams' goal.
 \item It is assumed that information is private, even among team members. Therefore, agents do not know other agents' utility functions, strategies, reservation utilities, or deadlines. On top of that, we also assume agents with bounded computational resources. Thus, we take a heuristic approach which seeks near optimal results while being computationally tractable.
 \item It is assumed that the team $A$ and the opponent $op$ communicate following an alternating bilateral protocol \cite{rubinstein82}. One of the two parties acts as the initiator, and it is entitled to propose the first offer. The other party receives the offer and can respond with two different actions: accept the offer (successful negotiation), or propose a counteroffer. If a counteroffer is proposed, the initiator party receives the offer and it can either accept the counteroffer or propose another offer, starting a new negotiation round. Depending on the intra-team strategy, one of the team members or a team mediator is responsible of the communications with the opponent. In this setting, the fact that one of the parties is a team remains unknown to the other party. 
 \item Additionally, it is also assumed that the negotiation is time-bounded, and each party has a private deadline $T_{A}$ (team deadline), $T_{op}$ (opponent deadline). When its deadline is achieved, the party leaves the negotiation and it is considered a failed negotiation. In the case of $T_{A}$, it is considered a joint deadline for all of the team members, who have agreed upon this deadline prior to the negotiation at hand. 
 \item The mediator, if present, is never a perfect mediator that aggregates the utility functions of all the team members. This assumption is taken due to the fact that, depending on the application, some team members may not be completely trustable and may attempt to exaggerate/change their preferences to manipulate the negotiation process. This mischievous behavior is easily carried out when aggregating utility functions. 
 \item The negotiation domain is comprised of $n$ real attributes whose domains can be scaled to $[0,1]$. Thus, the possible number of offers is $[0,1]^{n}$. In this domain, a complete offer is represented as $X=\{x_{1},x_{2},\dots,x_{n}\}$, where $x_{i}$ is a specific instantiation of attribute $i$. Additionally, we use the notation $X^{t}_{i\rightarrow j}$ to denote that offer $X$ was sent at round $t$ from party $i$ to party $j$. 
 \item Every agent $i$ (team member or opponent) has its preferences represented by means of linear additive utility functions in the form:
 \begin{equation}
   U_{i}(X)= w_{i,1}\;V_{i,1}(x_{i,1})+w_{i,2}\;V_{i,2}(x_{2})+...+w_{i,n}\;V_{i,n}(x_{n})
 \end{equation}
where $X$ is a complete offer, $x_{j}$, is the value given to the $j$-th attribute, $V_{i,j}(.)$ is the valuation function for attribute $j$ used by agent $i$ to normalize the attribute value to $[0,1]$, and $w_{i,j}$ is the weight/importance given by agent $i$ to attribute $j$ in the negotiation process. Several observations should be made regarding these utility functions: (i) weights are normalized so that $\sum_{j=1}^{n} w_{i,j}=1$; (ii) attributes are assumed to be independent from each other. Thus, the valuation of one of the attributes does not alter the others attributes' valuation; (iii) it is assumed that valuation functions are either monotonically increasing or monotonically decreasing. Moreover, we assume that team members share the same type of monotonic function (i.e., increasing or decreasing) for each $V_{i,j}(.)$. As for the opponent, it is assumed that the monotonic function for $V_{i,j}(.)$ is the opposite type to that of team members. It is reasonable to assume this model for valuation functions in e-commerce scenarios. Buyers usually share the same type of valuation function for attributes such as the price (monotonically decreasing), product quality (monotonically increasing), and the dispatch time (monotonically decreasing), whereas sellers usually use the opposite type of monotonic functions (monotonically increasing for price, monotonically decreasing for product quality, and monotonically increasing for dispatch time); (iv) attribute weights $w_{i,j}$ are different for team members. This way, we are able to represent the fact that some team members may be more interested in some attributes whereas other team members may be more interested in other attributes (e.g., some team members prefer price over quality, while others give a higher priority to the product quality). Obviously, the weights of the opponent's utility function may be different from those of team members; (v) since team members share the same type of monotonic function, if one of the team members increases its utility by increasing/decreasing one of the attribute values, the other team members will stay at the same utility level or they will also increase their utility. Thus, there is potential for cooperation among team members. 
\item The opponent has a reservation utility $RU_{op}$. Any offer whose utility is lower than $RU_{op}$ will be rejected. Each team member $a_{i}$ has a private reservation utility $RU_{a_{i}}$. This individual reservation utility is not shared among teammates. Therefore, a team member $a_{i}$ will reject any offer whose value is under $RU_{a_{i}}$. In this setting, reservation utilities represent the individual utility of each agent if the negotiation process fails. For the experiments, the reservation utilities are drawn from uniform distributions $RU_{op}=U[0,0.25]$ and $RU_{a_{i}}=U[0,0.25]$.

\end{itemize}

\subsection{Opponent Model}
\label{sec:opponent}
The opponent $op$ is modeled as a single agent for the sake of simplicity. We acknowledge that the other party may be another team (e.g., organization vs organization), but the focus of our study in this article, and the kind of applications in our mind, involve negotiations between one team and one single agent. For the opponent model we used well-known strategies in the agent negotiation literature:
\begin{itemize}
 \item The opponent uses a time-based tactic during the negotiation process. In multi-attribute, time-bounded negotiations, assuming that agents concede gradually to reach an agreement is usual. This is especially important when agents do not know other agents' preferences and strategies, since an exploration of the agreement space is required for discovering possible agreements.   A concession strategy typically starts by demanding the maximum aspiration and, as the negotiation process advances, the aspiration demanded tends to be lowered. The speed at which the strategy concedes is regulated by the concession speed parameter. In this article, we employed a time-based concession tactic inspired in tactics used by other authors \cite{faratin98,lai08}:
 \begin{equation}
 \label{time}
  s_{op}(t)=1-(1-RU_{op})(\frac{t}{T_{op}})^{\frac{1}{\beta_{op}}}
 \end{equation}
 where $t$ is the current negotiation round and $\beta_{op}$ is a parameter that governs the concession speed. On the one hand, when $\beta_{op}=1$ the concession is linear and each negotiation round the same amount of concession is performed, and when $\beta_{op}<1$ the concession is Boulware and very little is conceded at the start of the negotiation process but the agent concedes faster as the negotiation deadline approaches. On the other hand, when $\beta_{op}>1$ the tactic is conceder and the agents concede fast towards the reservation utility in the first rounds.
 \item The opponent uses an offer acceptance criterion $ac_{op}(.)$ during the negotiation process. It is formalized as follows:
 \begin{equation}
ac_{op}(X_{A\rightarrow op}^{t}) = \left\lbrace
	  \begin{array}{l l}
	     accept & \mbox{if } s_{op}(t+1) \leq U_{op}(X_{A\rightarrow op}^{t}) \\
	     reject & \mbox{otherwise}
	  \end{array}
	  \right.
 \end{equation}
 where $t$ is the current round, $X^{t}_{A \rightarrow op}$ is the offer received from the team, $U_{op}(.)$ is the utility function of the opponent, and $s_{op}(.)$ is the opponent concession strategy. Thus, an offer is accepted by $op$ if it reports a utility that is equal to or greater than the utility of the offer that $op$ would propose in the next round.
\item When the opponent has to propose a counteroffer at negotiation round $t$, it proposes an offer $X_{op\rightarrow A}^{t}$ whose utility is $U_{op}(X_{op\rightarrow A}^{t})=s_{op}(t)$. However, depending on the negotiation domain, there may be an infinite number of offers that comply with the previous condition. Similarity heuristics are a largely used family of heuristics in agent-based negotiation literature \cite{faratin02,lai08,sanchez-anguix10}. Thus, we assumed that the opponent attempts to propose the offer that is the most similar to the last offer received from the team, and whose utility is $U_{op}(X_{op\rightarrow A}^{t})=s_{op}(t)$. The Euclidean distance was used as similarity function.
\end{itemize}

\section{Intra-Team Strategies}
\label{sec:intra}
An intra-team strategy defines \textit{what} decisions have to be taken by a negotiation team, \textit{how} those decisions are taken, and \textit{when} those decisions are taken. In a bilateral negotiation process between a team and an opponent, the decisions that must be taken (\textit{what}) are which offers are sent to the opponent, and whether or not opponent offers are accepted. Given the fact that a negotiation team is formed by more than a single individual, decisions should take into account the interests of the team members. \textit{How} decisions are taken will determine the satisfaction level of the team with the final decision. In this article, most of the team interactions in intra-team strategies are carried out during the negotiation (\textit{when}). 

Next, we describe the four intra-team strategies that we propose: Representative (RE), Similarity Simple Voting (SSV), Similarity Borda Voting (SBV) and Full Unanimity Mediated (FUM). Each strategy is capable of guaranteeing a minimum level of unanimity regarding the offer sent to the opponent, and whether or not to accept the opponent's offer. Another difference between the four strategies is the presence of a mediator and its level of activity (none, coordination tasks, very active participation). Table \ref{tab:qualitative} captures the main qualitative differences between the four intra-team strategies according to the aforementioned criteria.

\begin{table}[ht]
\center
\scalebox{0.75}{
 \begin{tabular}{|l | l | l | l |}
\hline
\textbf{Intra-Team Strategy} & \textbf{Unanimity level} (How) & \textbf{Pre-negotiation?} (When) & \textbf{Mediated?} \\
\hline
RE & Unilateral & Minimum & No \\
\hline
SSV & Plurality/Majority & Minimum & Only coordination \\
\hline
SBV & Semi-Unanimity & Minimum & Only coordination \\
\hline
FUM & Unanimity & Information sharing & Very Active\\
\hline
  
 \end{tabular} 
}
\caption{A brief qualitative comparison among the four intra-team strategies presented in this article}
\label{tab:qualitative}
\end{table}  

\subsection{Representative (RE)}
\label{sec:re}
The Representative strategy (RE) is perhaps the simplest intra-team strategy. Basically, one of the team members is selected as representative $a_{re}$ for the team during the negotiation. This agent will act on behalf of the team during the negotiation, making it responsible of selecting which offers are sent to the opponent, and whether or not opponent's offers are accepted. The only communications are those carried out between the representative agent $a_{re}$ and the opponent $a_{op}$, and, therefore, this strategy is equivalent to a classic bilateral strategy. 

The representative agent negotiates according to its own utility function $U_{a_{re}}(.)$ since it does not know the utility function of the other participants. The two decisions that have to be taken during the negotiation are which offers are sent to the opponent, and whether or not the opponent's offer is accepted. 

\subsubsection{Offer proposal}
Being a time-bounded negotiation, the representative employs a time-based concession tactic $s_{a_{re}}(.)$ to negotiate with the opponent. It is based on a team deadline $T_{A}$ and a concession speed $\beta_{A}$, which have been agreed upon prior to the negotiation start: 
\begin{equation}
 s_{a_{re}}(t)=1-(1-RU_{a_{re}}) (\frac{t}{T_{A}})^{\frac{1}{\beta_{A}}}
\end{equation}
The concession strategy defines the aspiration level (utility demanded) by the agent at a specific round $t$. This utility is demanded from the point of view of the representative, and, so, any offer $X_{A\rightarrow op}^{t}$ proposed by $a_{re}$ at round $t$ will obey the following condition:
\begin{equation}
 U_{a_{re}}(X_{A\rightarrow op}^{t})= s_{a_{re}}(t)
\end{equation}
Since there is a large number of offers that may obey the equation above, we aimed to satisfy the opponent's preferences as much as possible. As in the case of the opponent strategy, the representative selects the offer that is the most similar to the previous offer received from the opponent using a similarity heuristic based on the Euclidean distance.
\begin{equation}
 X_{A\rightarrow op}^{t}=\underset{X | U_{a_{re}}(X)= s_{a_{re}}(t)}{max} Sim(X,X_{op\rightarrow A}^{t-1})
\end{equation}
\subsubsection{Offer acceptance}
A common acceptance criterion in time-bounded negotiations is that an opponent's offer is accepted if it reports a utility which is higher than or equal to the utility that is to be demanded in the next negotiation step. In the case of the representative, it will accept the opponent's offer $X_{op\rightarrow A}^{t}$ at round $t$ if it reports a utility $U_{a_{re}}(X_{op\rightarrow A}^{t})$ greater than or equal to $s_{a_{re}}(t+1)$. This can be formalized as follows:
 \begin{equation}
ac_{a_{re}}(X_{op\rightarrow A}^{t}) = \left\lbrace
	  \begin{array}{l l}
	     accept & \mbox{if } s_{a_{re}}(t+1) \leq U_{a_{re}}(X_{op\rightarrow A}^{t}) \\
	     reject & \mbox{otherwise}
	  \end{array}
	  \right.
 \end{equation}

\subsubsection{Unanimity Level}
It is clear that since the representative negotiates according to its own utility function and reservation utility, it cannot guarantee any kind of unanimity regarding team decisions. Decisions taken by the representative are acceptable to himself (1 agent), but nothing can be assured about the rest of team members. One could think that if no unanimity can be guaranteed, this strategy is not worth being used. However, when team members tend to be very similar this strategy is expected to yield acceptable results with communication costs equivalent to a bilateral negotiation process.

\subsection{Similarity Simple Voting (SSV)}
\label{sec:ssv}
The second intra-team strategy relies on a trusted mediator that helps team members to participate in the negotiation process. Its main tasks involve coordination of voting processes and communications with the opponent. It should be highlighted that the mediator communicates team's decisions to the opponent, and broadcasts opponent's decisions among team members. Thus, the fact that every team member participates in the negotiation process remains unknown for the opponent. As for intra-team communications, it should be noted that team members do not communicate among them, but they only communicate anonymously with the mediator. 

The decision rule used for voting processes is plurality/majority. More specifically, a plurality rule is used in the voting process employed to decide which offer is sent to the opponent, and a majority rule is used in the voting process employed to decide opponent's offer acceptance. A detailed view of the intra-team strategy can be observed in Algorithm \ref{alg:ssv}, which describes the whole process from the point of view of the mediator.
\subsubsection{Offer proposal}
Whenever a new offer has to be proposed to the opponent at round $t$, the mediator opens a call for proposals among team members. Each team member $a_{i}$ is allowed to communicate anonymously one offer $X_{a_{i}\rightarrow A}^{t}$ to be proposed to the opponent. Once every proposal has been gathered, the mediator opens a voting process where offers proposed $XT^{t}$ are made public among team members. Then, each agent $a_{i}$ anonymously sends a multi-vote $Vote_{a_{i}}$ to the mediator. A multi-vote gathers votes for every offer made public. We use the notation $Vote_{a_{i}}(j)$ to denote the vote given by agent $a_{i}$ to the offer $j$-th from $XT^{t}$, and $XT^{t}(j)$ as the $j$-th offer in $XT^{t}$. The votes can be either positive (1), if the offer $j$-th is acceptable for $a_{i}$ at round $t$, or negative (0), if the offer $j$-th is not acceptable for $a_{i}$ at round $t$. Once all votes have been gathered, the mediator sums up the number of positive votes and the most supported offer $X_{A\rightarrow op}^{t}$ is selected, made public among team members, and sent to the opponent. When a tie is produced, the tie-breaker rule consists in randomly selecting one of the most supported offers. The following Equation describes the selection rule of the previous mechanism:
\begin{equation}
X_{A\rightarrow op}^{t}=\underset{X_{j}\in XT^{t}}{\mbox{argmax}}\underset{a_{i}\in A}{\sum}Vote_{a_{i}}(j)
\label{eq:sum}
\end{equation}
The paragraph above describes the intra-team protocol followed by team members and mediator to determine which offer is sent to the opponent at round $t$. However, team members are faced with two decisions in this intra-team protocol: which offer should be proposed to the mediator during the call for proposals, and the acceptability of each offer proposed during the aforementioned process. We assume that, since the negotiation is time-bounded, team members follow a time-based concession tactic where the concession speed $\beta_{A}$ is common and agreed by teammates prior to the negotiation process:
\begin{equation}
 s_{a_{i}}(t)=1-(1-RU_{a_{i}})(\frac{t}{T_{A}})^{\frac{1}{\beta_{A}}}
\label{eq:time}
\end{equation}
For the first decision, proposing an offer to team members, the agent $a_{i}$ proposes an offer $X_{a_{i}\rightarrow A}^{t}$ whose utility is equal to $U_{a_{i}}(X_{a_{i}\rightarrow A}^{t})=s_{a_{i}}(t)$. Since there may be more than a single offer with such utility, the agent has to choose one of those offers. If the agent $a_{i}$ wants its offer $X_{a_{i}\rightarrow A}^{t}$ to be accepted it should maximize the probability of it being the most supported by team members and the probability of it being accepted by the opponent:
\begin{equation}
 X_{a_{i}\rightarrow A}^{t}= \underset{X|U_{a_{i}}(X)=s_{a_{i}}(t)}{argmax}p_{op}(X) \times p_{A}(X)
\label{eq:prob}
\end{equation}
where $p_{op}(X)$ is the probability for $X$ to be accepted by the opponent, and $p_{A}(X)$ is the probability for $X$ to be selected by team members. We incorporated agents with a similarity heuristic based on the Euclidean distance over attribute domains scaled to [0,1]. It takes into account the last offer proposed by the opponent $X_{op\rightarrow A}^{t-1}$ and the offer sent by team members in the previous negotiation round $X_{A\rightarrow op}^{t-1}$. The most similar an offer is to $X_{op\rightarrow A}^{t-1}$, the more probable it is for the offer to be accepted by the opponent. Analogously, the most similar an offer is to $X_{A\rightarrow op}^{t-1}$, the more probable it is for the offer to be the most supported option in the voting process and, therefore, to be sent to the opponent. Thus, Equation \ref{eq:prob} can be approximated by similarity heuristics as follows:
\begin{equation}
\begin{array}{l l}
 X_{a_{i}\rightarrow A}^{t}= &  \underset{X|U_{a_{i}}(X)=s_{a_{i}}(t)}{argmax}p_{op}(X) \times p_{A}(X)  \approx\\
 & \underset{X|U_{a_{i}}(X)=s_{a_{i}}(t)}{argmax} Sim(X,X_{op\rightarrow A}^{t-1}) \times Sim(X,X_{A\rightarrow op}^{t-1})
\end{array}
\label{eq:similarity}
\end{equation}
Finally, for determining the acceptability of offers proposed by team members at round $t$, we used a rational criterion so that an agent $a_{i}$ emits a positive vote $Vote_{a_{i}}(j)=1$ for the $j$-th offer if it reports a utility that is greater or equal than the utility marked by the concession strategy $s_{a_{i}}(t)$. Otherwise, the offer is not supported and a negative vote is emitted. This process can be formalized as:
\begin{equation}
Vote_{a_{i}}(j) = \left\lbrace
	  \begin{array}{l l}
	     1 & \mbox{\;\;if } U_{a_{i}}(XT^{t}(j)) \geq s_{a_{i}}(t) \\
	     0 & \mbox{\;\;otherwise}
	  \end{array}
	  \right.
\end{equation}

\subsubsection{Offer acceptance}
Whenever the mediator receives an offer $X_{op\rightarrow A}^{t}$ from the opponent at round $t$, it broadcasts the offer among team members. Then, the mediator opens up a majority voting process where each agent $a_{i}$ states whether or not the opponent's offer is acceptable $ac_{a_{i}}(X_{op\rightarrow A}^{t})$ (1 for accept, 0 for reject). The mediator counts the number of acceptances, and if the offers is supported by the majority ($>\frac{|A|}{2}$) then it is accepted by the team. Otherwise, the offer is rejected. If the number of team members is even and a tie has been produced, a random decision is taken by the mediator. This mechanism can be described as follows:
\begin{equation}
ac_{A}(X_{op\rightarrow A}^{t}) = \left\lbrace
	  \begin{array}{l l}
	     \mbox{accept} & \mbox{\;\;if } \underset{a_{i}\in A}{\sum}ac_{a_{i}}(X_{op\rightarrow A}^{t}) > \frac{|A|}{2}\\
	     \mbox{reject} & \mbox{\;\;if } \underset{a_{i}\in A}{\sum}ac_{a_{i}}(X_{op\rightarrow A}^{t}) < \frac{|A|}{2} \\
	     \mbox{random} & \mbox{\;\;otherwise}
	  \end{array}
	  \right.
\end{equation}

How team members $a_{i}$ decide the acceptability of the opponent's offer $ac_{a_{i}}(X_{op\rightarrow A}^{t})$ follows the rational mechanism that we have employed so far. Basically, the offer is acceptable (1) if it yields a utility which is greater than or equal to the utility demanded by the concession strategy in the next negotiation round $s_{a_{i}}(t+1)$. Otherwise, the offer is not considered acceptable. The following Equation formalizes the acceptance criterion:
\begin{equation}
ac_{a_{i}}(X_{op\rightarrow A}^{t}) = \left\lbrace
	  \begin{array}{l l}
	     \mbox{1} & \mbox{\;\;if } U_{a_{i}}(X_{op\rightarrow A}^{t}) \geq s_{a_{i}}(t+1)\\
	     \mbox{0} & \mbox{\;\;otherwise}
	  \end{array}
	  \right.
\end{equation}
\begin{algorithm}[H]
\scriptsize
$t=0$\;
\While{$t\leq T_{A}$}{
Send (Call For Proposals $\longrightarrow$ $A$)\;
$XT^{t}=\emptyset$\;
\ForEach{$a_{i} \in A$}{
Receive ($X_{a_{i}\rightarrow A}^{t}$ $\longleftarrow$ $a_{i}$)\;
$XT^{t}=XT^{t}\bigcup X_{a_{i}\rightarrow A}^{t}$\;
}
Send (Open Voting $XT^{t}$ $\longrightarrow$ $A$)\;
\ForEach{$a_{i} \in A$}
{
Receive ($Vote_{a_{i}}$ $\longleftarrow$ $a_{i}$)\;
}
$X_{A\rightarrow op}^{t}=\underset{X_{j}\in XT^{t}}{\mbox{argmax}}\underset{a_{i}\in A}{\sum}Vote_{a_{i}}(j)$\;
Send ($X_{A\rightarrow op}^{t}$ $\longrightarrow$ $op,A$)\;
Receive ($X_{op\rightarrow A}^{t}$ $\longleftarrow$ $op$)\;
\If{$X_{op\rightarrow A}^{t}=$ Withdraw}{Send (Opponent Withdraw $\longrightarrow$ $A$)\;Return Failure\; }
\ElseIf{$X_{op\rightarrow A}^{t}=$ Accept}{Send (Offer Accepted $\longrightarrow$ $A$)\;Return Success\; }
\Else{
Send (Open Voting $X_{op\rightarrow A}^{t}$ $\longrightarrow$ $A$)\;
\ForEach{$a_{i} \in A$}{
Receive ($ac_{a_{i}}(X_{op\rightarrow A}^{t})$ $\longleftarrow$ $a_{i}$)\;
}

 \If{$ac_{A}(X_{op\rightarrow A}^{t})$= accept}{Send (Accept $\longrightarrow$ $op,A$)\;Return Success\;}
 \Else{Send (Opponent Offer Rejected $\longrightarrow$ $A$)\;}
}
  $t=t+1$\;
}
Send (Withdraw $\longrightarrow$ $op,A$)\;
Return Failure\;

\caption{Pseudo-code algorithm for the mediator in the Similarity Simple Voting intra-team strategy. Messages are represented as (\textit{Body} \textit{direction} \textit{agents}). Therefore, (Accept $\longrightarrow$ $op$) means that the agent sends an accept message to $op$, whereas (Reject $\longleftarrow$ $op$) describes a message from $op$ with the content ``Reject''. }
\label{alg:ssv}
\end{algorithm}

\subsubsection{Unanimity Level}
The proposed method is capable of guaranteeing team decisions that are supported by a plurality/majority of the participants. More specifically, plurality is assured in case of the offer proposed to the opponent, and majority is assured when deciding opponent's offer acceptance. Exceptions for this minimum level of team unanimity are ties. For instance, the most extreme case is when team members propose offers to the team, but they only support their own offers. In that case, each proposal sums up exactly 1 positive vote and there is not a clear plurality winner.

\subsection{Similarity Borda Voting (SBV)}
\label{sec:sbv}

SSV is capable of assuring majority and plurality decisions within the team. However, some scenarios may need of intra-team strategies that ensure higher levels of unanimity. SBV and FUM (described later) are designed to solve this problem. The basic structure of SBV remains the same than in SSV, but the voting rules employed are different. More specifically, when each team member votes team proposals, borda count is employed to determine the winner, and a unanimity rule is used to determine opponent's offer acceptance. Next, we briefly describe the aspects which make SBV different to SSV.

\subsubsection{Offer proposal}
As in SSV, when the team has to propose an offer to the opponent, the mediator opens a call for proposals where each team member can propose an offer to the mediator. Then, once every offer has been gathered, the mediator makes public the offers proposed to the team members and a voting process starts. The main difference between both intra-team strategies resides in the fact that team members vote according to a Borda count rule \cite{nurmi10}. Basically, each team member $a_{i}$ ranks the proposals $XT^{t}$ in ascending order according to its own utility function $U_{a_{i}}(.)$. We denote as $rank_{a_{i}}(XT^{t})$ the ascending rank according to $a_{i}$'s utility function, and $Position(X,rank_{a_{i}}(XT^{t}))$ as the position (1 to $|XT^{t}|$) that the offer $X$ occupies in a ranked list. The vote emitted by $a_{i}$ for offer $j$-th in $XT^{t}$ is the position occupied by such offer in the ranked list minus one unit:
\begin{equation}
Vote_{a_{i}}(j) = Position(XT^{t}(j),rank_{a_{i}}(XT^{t}))-1
\end{equation}
Numerical votes for each offer are summed up by the mediator, who finally selects the offer that received the highest sum of scores from the team members (see Equation \ref{eq:sum}). It should be highlighted that the similarity heuristic employed by team members is the same than the one employed in SSV.
\subsubsection{Offer acceptance}
As for the offer acceptance, the only difference remains in the rule used by the mediator. The opponent's offer is accepted only if it is acceptable for all the team members. The rationale $ac_{a_{i}}(X_{op\rightarrow A}^{t})$ used by team members to determine if an offer is acceptable at round $t$ is equivalent to the one used in SSV. Thus, the offer acceptance mechanism can be formalized as follows:
\begin{equation}
ac_{A}(X_{op\rightarrow A}^{t}) = \left\lbrace
	  \begin{array}{l l}
	     \mbox{accept} & \mbox{\;\;if } \underset{a_{i}\in A}{\sum}ac_{a_{i}}(X_{op\rightarrow A}^{t}) = |A| \\
	     \mbox{reject} & \mbox{\;\;otherwise }\\
	  \end{array}
	  \right.
\end{equation}
\subsubsection{Unanimity Level}
When describing the minimum unanimity level guaranteed by SBV, we mentioned the term semi-unanimity. It is clear that if an opponent offer is accepted by the team, it is acceptable for every team member due to the unanimity ruled employed. However, such unanimity is not guaranteed regarding the team decision on which offer is sent to the opponent. Borda count is generally referred as a method that selects broadly accepted options as winners instead of the majority/plurality option (e.g., avoid the tyranny of the majority). In this sense, Borda count entails some degree of unanimity. Nevertheless, the specific degree of unanimity that Borda assures is difficult to determine in our negotiation scenario.

\subsection{Full Unanimity Mediated (FUM)}
\label{sec:fum}
The last intra-team strategy, Full Unanimity Mediated (FUM), seeks to reach unanimity regarding all team decisions. In fact, every team decision taken (i.e., offer acceptance, offer proposal) following this intra-team strategy entails unanimity at each round $t$ of the negotiation process. However, the type of mediator required for FUM is more sophisticated than in the rest of strategies presented in this article. It requires that the mediator participates in a pre-negotiation process where team members hand over decision rights over attributes that are not interesting for them. Additionally, the team mediator needs to be able to infer attributes' importance for the opponent. Finally, it also needs to coordinate unanimity voting processes, and an iterated building process that constructs the offers sent to the opponent. A complete view of the pseudo-algorithm carried out by the mediator can be observed in Algorithm \ref{alg:fum}.

\begin{algorithm}
\scriptsize
/*Pre-negotiation: information sharing*/\;
Send (Ask for $NI_{a_{i}}$ $\longrightarrow$ $A$)\;
\lForEach{$a_{i} \in A$}{ 
Receive ($NI_{a_{i}}$ $\longleftarrow$ $a_{i}$)\;}
$t=0$\;
\While{$t\leq T_{A}$}{
/*Offer proposal starts*/\;
$agenda=build\_agenda()$\;
$A'=A;\;\;X_{A\rightarrow op}^{'t}=\emptyset$\;
/*Attributes not interesting for every team member are maximized for the opponent*/\;
\ForEach{$j \in \overset{M}{\underset{i=1}{\bigcap}}NI_{a_{i}} $}{
$x_{j}=maximize\_for\_opponent(j)$\;
$X_{A\rightarrow op}^{'t}=X_{A\rightarrow op}^{'t} \bigcup \{x_{j}\}$\;
}

/*The offer is built attribute per attribute*/\;
\ForEach{$j \in agenda \wedge attribute\_not\_set(j)$}{
/*Ask about the value needed of attribute j*/\;
Send (Needed value $j$, given $X_{A\rightarrow op}^{'t}$ $\longrightarrow$ $a_{i}|j\notin NI_{a_{i}}\wedge a_{i}\in A'$)\;
Receive ($x_{a_{i},j}$ $\longleftarrow$ $a_{i}|j\notin NI_{a_{i}}\wedge a_{i}\in A'$)\;
\lIf{$monotonically\_increasing(j)$}{
$x_{j}=\underset{a_{i}\in A}{max}\;\;x_{a_{i},j}$\;}
\lElse{$x_{j}=\underset{a_{i}\in A}{min}\;\;x_{a_{i},j}$\;}
$X_{A\rightarrow op}^{'t}=X_{A\rightarrow op}^{'t} \bigcup \{x_{j}\}$\;
/*Ask about the acceptability of the partial offer*/\;
Send (Acceptable $X_{A\rightarrow op}^{'t}$? $\longrightarrow$ $a_{i}|a_{i}\in A'$)\;
\ForEach{$a_{i}\in A'$}{
Receive ($ac'_{a_{i}}(X_{A\rightarrow op}^{'t})$ $\longleftarrow$ $a_{i}$)\;
\lIf{$ac'_{a_{i}}(X_{A\rightarrow op}^{'t})=true$}{$A'=A'-\{a_{i}\}$\;}
}
/*If no more active agents, we do not ask team members anymore*/\;
\lIf{$A'=\emptyset$}{break;}
}
/*Any attribute not set yet, is maximized for the opponent*/\;
\ForEach{$j \in agenda \wedge attribute\_not\_set(j)$}
{
$x_{j}=maximize\_for\_opponent(j)$\;
$X_{A\rightarrow op}^{'t}=X_{A\rightarrow op}^{'t} \bigcup \{x_{j}\}$\;
}
$X_{A\rightarrow op}^{t}=X_{A\rightarrow op}^{'t}$\;
Send ($X_{A\rightarrow op}^{t}$ $\longrightarrow$ $op,A$)\;

/*Opponent offer acceptance starts*/\;
Receive ($X_{op\rightarrow A}^{t}$ $\longleftarrow$ $op$)\;
\uIf{$X_{op\rightarrow A}^{t}=$ Withdraw}{Send (Opponent Withdraw $\longrightarrow$ $A$)\;Return Failure\; }
\uElseIf{$X_{op\rightarrow A}^{t}=$ Accept}{Send (Offer Accepted $\longrightarrow$ $A$)\;Return Success\; }
\Else{
Send (Open Voting $X_{op\rightarrow A}^{t}$ $\longrightarrow$ $A$)\;
\lForEach{$a_{i} \in A$}{
Receive ($ac_{a_{i}}(X_{op\rightarrow A}^{t})$ $\longleftarrow$ $a_{i}$)\;
}

 \uIf{$ac_{A}(X_{op\rightarrow A}^{t})$= accept}{Send (Accept $\longrightarrow$ $op,A$)\;Return Success\;}
 \Else{Send (Opponent Offer Rejected $\longrightarrow$ $A$)\;}
}
  $t=t+1$\;
}
Send (Withdraw $\longrightarrow$ $op,A$)\;
Return Failure\;

\caption{Pseudo-code algorithm for the mediator in the FUM intra-team strategy. }
\label{alg:fum}
\end{algorithm}

\subsubsection{Pre-negotiation: information sharing}
During the pre-negotiation, team members are allowed to hand over decision rights over some attributes that they do not consider interesting. The iterated offer building process relies on a mechanism which sets attributes' values one-per-one according to team members' will. When an agent hands over decision rights on an attribute, it does not participate in the setting of such attribute. All the communications in the pre-negotiation are private with the mediator, who asks each team member regarding the set of attributes which it is willing to hand over. The rationale behind the idea of handing over decision rights is that conflict may be reduced, and, so, the chances to build a more likeable offer for the opponent are increased while maintaining a good quality for one's own utility function. The fact that some attributes may yield little or no importance at all for some team members is also feasible in a team setting, since some of these attributes may have been introduced to satisfy the interests of a subgroup of team members.

The pre-negotiation protocol goes as follows. First, the mediator opens a call for decision rights, where each team member $a_{i}$ is allowed to send (to the mediator) a set of negotiation attributes $NI_{a_{i}}$, whose decision rights are handed over by $a_{i}$. Once all the responses have been gathered, the mediator keeps track of those attributes that are not interesting for each agent $NI_{a_{i}}$, and those attributes that are not interesting for all team members $\overset{M}{\underset{i=1}{\bigcap}}NI_{a_{i}}$. Once this process has finished, the team and the mediator are ready to start the negotiation process.

Of course, the set of attributes handed over by each team member is not controllable by the mediator. It depends on the behavior of each agent. In our model, the set of attributes handed over by each agent depends on a private parameter $\epsilon_{a_{i}}$. The value of such parameter is related to the weight of the different negotiation attributes in one's own utility function. More precisely, if $\epsilon_{a_{i}}=0$, then the agent is only willing to hand over the decision rights over those attributes that are not interesting for himself (i.e., weight equal to zero in the utility function). When $\epsilon_{a_{i}}=1$, the agent is willing to hand over decision rights over every attribute in the negotiation. In general,  the agent is willing to hand over decision rights over attributes whose sum of weight in the utility function is equal to or lower than $\epsilon_{a_{i}}$: 
\begin{equation}
\label{eq_team_tolerance}
 \underset{j \in NI_{a_{i}}}{\sum}w_{a_{i},j} \leq \epsilon_{a_{i}}
\end{equation}

 Given a certain $\epsilon_{a_{i}}$, a reasonable heuristic is to assume that the agent is willing to concede as many decision rights as possible since this will enhance the possibility of finding an agreement with the opponent. Hence, each team member $a_{i}$ chooses the largest possible set $NI_{a_{i}}$ that fulfills Eq. \ref{eq_team_tolerance}. A simple algorithm that solves this problem is ordering the negotiation attributes in ascending order by weight in the utility function. The set $NI_{a_{i}}$ starts empty, and, then, the array of ordered attributes is followed. If the attribute weight plus the weights of those attributes already in $NI_{a_{i}}$ exceeds $\epsilon_{a_{i}}$, then the search stops. Otherwise, the attributes is added to $NI_{a_{i}}$ and the algorithm continues with the next attribute. Our initial experiments with FUM \cite{sanchez-anguix12} suggested that team members should set its private $\epsilon_{a_{i}}$ to 0 and hand over decision rights only over those attributes that are not interesting at all. Therefore, in the experimental setting, we use $\epsilon_{a_{i}}=0$.

\subsubsection{Negotiation: observing opponent's concessions and building an attribute agenda}
Once the negotiation starts, the mediator attempts to guess a ranking of attributes according to the opponent's preferences. This ranking is used to build an agenda of attributes, which is used in the iterated offer building process. The idea behind the agenda is attempting to satisfy team members as much as possible with those attributes that are less important for the opponent. This way, team members may reach their desired aspiration level with those attributes less interesting for the opponent, and use the rest of attributes to make the offer as satisfactory as possible for the opponent. The only information available for the mediator regarding the opponent's preferences are the offers received. Thus, the mediator has to infer a ranking of attributes according to that information. A possible heuristic is assuming that agents usually concede less in important attributes and greater concessions are performed in lesser important attributes at the first rounds of the negotiation.

Our proposed heuristic assumes that the mediator observes opponent's offers for the first $k$ interactions. In our experiments, the value of this parameter was set to $k=\lfloor \frac{T_{A}}{4} \rfloor$ \cite{sanchez-anguix12}. Then, it calculates the concession performed in each attribute. Since our model assumes that the opponent's utility function employs the opposite type of valuation function than team members for each attribute, it is relatively easy to calculate the amount of concession performed at each attribute. For instance, if the opponent is a seller, it is reasonable to assume that its valuation functions is monotonically increasing (e.g., higher prices report higher utilities) and, thus, any value below the maximum price can be considered a concession with respect to the maximum price. Therefore, the relative concession can be calculated in each attribute. For each attribute $j$, we calculate the total amount of relative concession $C_{j}$ in the first $k$ offers:
\begin{equation}
 C_{j}=\overset{k-1}{\underset{t=0}{\sum}} \frac{|X(j)_{op\rightarrow A}^{t}-best\_value(j)|}{max\_value(j)-min\_value(j)}
\label{eq:con}
\end{equation}
where $X(j)_{op\rightarrow A}^{t}$ it the value of attribute $j$ in the offer $X_{op\rightarrow A}^{t}$, $best\_value(j)$ is the best possible value for the opponent in attribute $j$, and $max\_value(j)$ and $min\_value(j)$ are the maximum and minimum value of the attribute in the negotiation domain. The inner part of the summatory determines the relative concession on attribute $j$ in the offer received at interaction/round $t$. So, the summatory counts the total relative concession for attribute $j$ in the first $k$ offers. The heuristic is that attributes that score lower in Equation \ref{eq:con} are usually those more important for the opponent, whilst those attributes scoring higher in Equation \ref{eq:con} are those less important for the opponent. Based on the available information (i.e., number of rounds up to $k$), the mediator builds an agenda of attributes according to the scores of $C_{j}$ in descending order. This way, lesser important attributes for the opponent are first in the agenda.

\subsubsection{Negotiation: Offer proposal}
In order to determine which offer is sent to opponent, the mediator governs an iterated building process. The aim of this iterated process is building an offer, attribute per attribute, so that the offer sent to the opponent is acceptable for every team member. The order in which the attributes are adjusted is determined by the agenda built by the mediator. The first attribute in the agenda is the one considered less important for the opponent, the second attribute is the next lesser important attribute for the opponent, and so forth. Thus, the first attributes set are those less important for the opponent. The heuristic used by this iterated building process is attempting to satisfy team members' demands with those attributes that are less important for the opponent, and demand as less as possible from those attributes that are the most important for the opponent. Briefly, the iterated building process goes as follows.

\begin{enumerate}
 \item The agenda of attributes $agenda$ is built by the mediator according to the available information. The first attribute in the agenda is the one guessed as the less important attribute for the opponent.
 \item When the iterated process starts, every team member is considered an active member ($a_{i} \in A'$) in the construction process.
  \item The initial partial offer $X_{A\rightarrow op}^{'t}$ starts as an offer whose attributes have not been set.
 \item The mediator checks those attributes that are not interesting for every team member $\overset{M}{\underset{i=1}{\bigcap}}NI_{a_{i}}$. These attributes are maximized according to the opponent's preferences (i.e., if the price was one of these attributes, it would be maximized for the opponent, thus, acquiring its minimum value). The partial offer $X_{A\rightarrow op}^{'t}$ is updated with the new attributes' values.
 \item The next attribute $j$ in the agenda is selected. Those team members active in the construction process ($a_{i} \in A'$) and interested in $j$ ($j \notin NI_{a_{i}}$) are asked by the mediator to submit the value $x_{a_{i},j}$ needed of attribute $j$ to get as close as possible to their aspiration levels.
 \item The values $x_{a_{i},j}$ gathered from team members are aggregated. If the assumed valuation function is monotonically increasing, then the $max$ operator is used to aggregate the values and obtain the final value for the attribute $x_{j}$. Otherwise, if the assumed valuation function is monotonically decreasing, then the $min$ operator is used to aggregate the values and obtain $x_{j}$.
 \item $x_{j}$ is set in $X_{A\rightarrow op}^{'t}$ and the new partial offer is broadcasted among team members. Every team member that is active in the construction phase is asked if the current partial offer satisfies its current demands.
 \item Every response is gathered by the mediator. Those agents that answered positively are removed from the list of active agents. If there are still active agents, the mediator goes back to 5.
 \item When every team member has been satisfied by the partial offer $X_{A\rightarrow op}^{'t}$, if there are still attributes that have not been set, those attributes are maximized according to the opponent's preferences. Then, a final offer $X_{A\rightarrow op}^{t}$ is obtained, made public among team members, and sent to the opponent.
\end{enumerate}

In the protocol described above, team members are asked to submit a value for attributes in which they are interested, and to determine whether or not the partial offer satisfies their needs. In both cases, as in previous strategies, we have assumed that team members follow time-based concession tactics similar to the one described in Equation \ref{eq:time}, where $\beta_{A}$ has been agreed upon by team members prior to the negotiation process. However, since team members may have handed over some decision rights, it is not possible for agents to demand the maximum utility. The value $\epsilon_{a_{i}}$ has to be subtracted from the maximum utility. Therefore, the concession strategy $s_{a_{i}}(t)$, which determines the level of demand at each negotiation round, can be formalized as:
\begin{equation}
 s_{a_{i}}(t)=(1-\epsilon_{a_{i}})-(1-\epsilon_{a_{i}}-RU_{a_{i}})(\frac{t}{T_{A}})^{\frac{1}{\beta_{A}}}
\end{equation}

When team members are asked about a value for $j$, each team member communicates anonymously the value $x_{a_{i},j}$. The value communicated is the one that gets as close as possible to its desired aspiration level $s_{a_{i}}(t)$ at round $t$. Taking the linear additive utility function formula, this can be calculated as:
\begin{equation}
\label{eq:bid1}
 x_{a_{i},j}= \underset{x\in [0,1]}{\mbox{argmin }} (s_{a_{i}}(t)-U_{a_{i}}(X_{A\rightarrow op}^{'t})-w_{a_{i},j}V_{a_{i},j}(x))
\end{equation}
where $s_{a_{i}}(t)$ is the utility demanded by the agent $a_{i}$ at round $t$, $U_{a_{i}}(X_{A\rightarrow op}^{'t})$ is the utility reported by the current partial offer, and $w_{a_{i},j}V_{a_{i},j}(x)$ is the weighted utility reported by the value demanded by the agent. Since the value demanded looks to be as close as possible to the utility necessary to get to the current aspiration, the function is minimized. However, the following constraint is fulfilled by team members in order to avoid surpassing the utility demanded:
\begin{equation}
\label{eq:bid2}
 s_{a_{i}}(t)-U_{a_{i}}(X_{A\rightarrow op}^{'t})-w_{a_{i},j}V_{a_{i},j}(x_{a_{i},j})\geq 0
\end{equation}

As for determining when a partial offer is acceptable, team members follow a similar criterion to the method proposed in other intra-team strategies. Basically, a partial offer is acceptable for an agent $a_{i}$ if it reports a utility that is greater than or equal to the aspiration level marked by its concession strategy:
\begin{equation}
\label{eq:accept}
 ac'_{a_{i}}(X_{A\rightarrow op}^{'t}) = \left\lbrace
	  \begin{array}{l l}
	     true & \mbox{if } U_{a_{i}}(X_{A\rightarrow op}^{'t}) \geq s_{a_{i}}(t) \\
	     false & \mbox{otherwise}
	  \end{array}
	  \right.
\end{equation}
where $true$ indicates that the partial offer is acceptable at its current state for agent $a_{i}$, and $false$ indicates the opposite.

\subsubsection{Negotiation: Offer acceptance}
Since this strategy looks for unanimity regarding team decisions, we employed the same mechanism employed in SBV for determining whether or not an opponent offer is acceptable. When the mediator receives the opponent's offer $X_{op\rightarrow A}^{t}$, the offer is publicly announced to all of the team members. Then, the mediator opens a private voting process where each team member $a_{i}$ should specify whether or not it supports acceptance of the opponent's offer $ac_{a_{i}}(X_{op\rightarrow A}^{t})$. The mediator counts the number of positive votes. The offer is accepted if the number of positive votes is equal to the number of team members. Otherwise, the offer is rejected.

Similarly to SBV, an opponent offer is acceptable for a team member at round $t$ if it reports a utility that is greater than or equal to the aspiration level marked by the concession strategy in the next round:
\begin{equation}
\label{op_ac}
ac_{a_{i}}(X_{op\rightarrow A}^{t}) = \left\lbrace
	  \begin{array}{l l}
	     true & \mbox{if } s_{a_{i}}(t+1) \leq U_{a_{i}}(X_{op\rightarrow A}^{t}) \\
	     false & \mbox{otherwise}
	  \end{array}
	  \right.
\end{equation}
where $true$ means that the agent supports the opponent's offer, $false$ has the opposite meaning, and $s_{a_{i}}(.)$ is the concession strategy employed by agent $a_{i}$ to calculate the aspiration level at each negotiation round $t$.

\subsubsection{Unanimity Level}
As stated in the introduction of this section, this strategy is capable of guaranteeing unanimity regarding team decisions. How unanimity is guaranteed in the offer acceptance phase is clear, since a voting process with unanimity rule is employed. In \cite{sanchez-anguix12} we showed how unanimity is also guaranteed in the offer sent to the opponent. More specifically, the strategy is capable of guaranteeing a strict unanimity: for any team member $a_{i}$, the offer sent to the opponent reports a utility that is greater than or equal to its aspiration level $s_{a_{i}}(t)$. This is possible thanks to the iterated building process and the assumptions in team members' utility functions. Since team members share the same type of monotonic valuation functions, the use of the max/min operator (max for monotonically increasing valuation functions, min for monotonically decreasing functions) ensures that for each attribute, each team member either gets exactly the value demanded for the attribute or it gets a value that reports a utility greater than or equal to the utility they demanded for the attribute. Hence, when team members demand the exact value needed to get as close as possible to their desired utility level, they will always get the same or greater utility than the one they actually demanded. Thus, in the end, the offer will yield a utility that is equal to or greater than their aspiration levels at round $t$. 

\section{Experimental Analysis}
\label{sec:exp}
In this section we study how the four intra-team strategies presented in this article perform under different environmental conditions. First, we introduce the negotiation case employed for our experiments. Then, the environmental conditions and performance measures studied are introduced and explained to the reader. Finally, we describe the experiments carried out, and we analyze the results provided by each intra-team strategy.
\subsection{Negotiation Case: Group Booking}

 The negotiation case employed for our experiments is based on a group booking negotiation with a hotel, which also illustrates the types of applications that can be built using the intra-team strategies proposed in this article. In this scenario, a group of friends who have decided to spend their holidays together has to book accommodation for their stay. Their destination is Rome, and they want to spend a whole week. Each friend is represented by his/her electronic agent, who acts semi-automatically on behalf of its user. This agent has previously elicited the preferences of its user regarding booking conditions. Each group member has different preferences regarding possible booking conditions. Thus, the final agreement with the hotel should satisfy every friend as much as possible. The group of agents engages in a negotiation with a well-known hotel in their city of destination, which is also represented by an electronic agent. During the pre-negotiation, both parties have decided to negotiate the following issues:
 \begin{itemize}
  \item Price per person ($pp$): The price per person is the amount of money that each friend will pay to the hotel for the accommodation service. The issue domain goes from 210\$, which is the minimum rate (30\$ per night), to 700\$, which is the maximum rate (100\$ per night). A realistic assumption in the group of friends is that friends prefer to pay lower prices to higher prices (i.e., monotonically decreasing valuation function), whereas the seller prefers to charge higher prices to lower prices (i.e., monotonically increasing valuation function).
  \item Cancellation fee per person ($cf$): When a booking is cancelled, the hotel charges a fee to compensate for losses. The issue domain goes from 0\$ (no cancellation fee) to 150\$. A realistic assumption in the group of friends is that friends prefer to pay lower prices to higher prices (i.e., monotonically decreasing valuation function), whereas the seller prefers to charge higher prices to lower prices (i.e., monotonically increasing valuation function).
  \item Full payment deadline ($pd$): The full payment deadline indicates when the group of friends has to pay the full price booking in order to confirm their reservation. The domain goes from ``Today''=0 days (the date time when the final agreement has been signed) to ``Departure Date''=30 days, which indicates that the team should only pay when leaving the hotel. A realistic assumption in the group of friends is that friends prefer to pay as late as possible (i.e., monotonically increasing valuation function), whereas the seller prefers to charge as soon as possible (i.e., monotonically decreasing valuation function).
  \item Discount in bar ($db$): As a token of respect for good clients, the hotel offers nice discounts at the hotel bar. The issue domain goes from 0\% (no discount) to 20\%. A realistic assumption in the group of friends is that friends prefer higher discounts to lower discounts (i.e., monotonically increasing valuation function), whereas the seller prefers to offer lower discounts prices to higher discounts (i.e., monotonically decreasing valuation function).
 \end{itemize}

\subsection{Negotiation Environment Conditions \& Team Performance}
We consider that the negotiation environment plays a very important part in team dynamics. It may not be the same using a representative approach in a setting where all of the team members' preferences are very similar than a setting where team members' preferences are exactly the opposite. Since conditions of the negotiation environment highly vary depending on the application domain, we decided to focus on those general conditions that are present in almost every negotiation scenario involving negotiation teams: opponent deadline, team deadline, team members' preference similarity, opponent concession speed, and team size.

Regarding team performance, it is also acknowledged that there are several well known social welfare measures to assess the quality of decisions in a society. A negotiation team can be considered a small society, and, thus, social welfare measures can also be considered appropriate measures for measuring negotiation teams' performance. More specifically, we study the impact of the negotiation environment on the minimum utility of team members (i.e., egalitarian social welfare \cite{chevaleyre06}), and the average utility of team members (i.e., a special case of ordered weighted averaging \cite{chevaleyre06}). However, we do not only restrain our analysis to social welfare measures. Computational measures like the number of negotiation rounds are also analyzed for all of the intra-team strategies.
\subsubsection{Environment Condition: Opponent Deadline Length}
One of the issues that can affect the negotiation process is the number of interactions that the opponent has until he decides that negotiating is no longer worthy, namely opponent deadline $T_{op}$. We partitioned the opponent negotiation deadline in three different classes: short deadline $T_{op}=U[5,10]=S$\footnote{U[5,10] is a uniform distribution from 5 to 10}, medium deadline $T_{op}=U[11,29]=M$, and long deadline $T_{op}=U[30,60]=L$.

\subsubsection{Environment Condition: Team Deadline Length}
Similarly, the maximum number of rounds that the team has to negotiate also may impact the performance of the different intra-team strategies. As in the case of the opponent deadline, we partitioned the team deadline in three different classes: short deadline $T_{A}=U[5,10]=S$, medium deadline $T_{A}=U[11,29]=M$, and long deadline $T_{A}=U[30,60]=L$.

\subsubsection{Environment Condition: Team Similarity}

25 different linear utility functions were randomly generated. These utility functions represented the preferences of potential team members. 25 linear utility functions were generated to represent the preferences of opponents. These utility functions were generated by taking potential teammates' utility functions and reversing the type of $V_{i}(.)$. 

In order to determine the preference diversity in a team, we decided to compare team members' utility functions. We introduce a dissimilarity measure based on the utility difference between offers. The dissimilarity between two teammates can be measured as follows: 
 \begin{equation}
 D(U_{a_{i}}(.),U_{a_{j}}(.))= \frac{\underset{\forall X \in [0,1]^{n}}{\sum} | U_{a_{i}}(X)-U_{a_{j}}(X) |}{|X \in [0,1]^{n}|}
 \end{equation}
 If the dissimilarity between two team members is to be measured exactly, it needs to sample all of the possible offers. However, this is not feasible in the current domain where there is an infinite number of offers. Therefore, we limited the number of sampled offers to 1000 per dissimilarity measure. Due to the fact that a team is composed by more than two members, it is necessary to provide a team dissimilarity measure. We define the team dissimilarity measure as the average of the dissimilarity between all of the possible pairs of teammates. 

 For all of the teams that had been generated, we measured their dissimilarity and calculated the dissimilarity mean $\bar{dt}$ and standard deviation $\sigma$. We used this information to divide the spectrum of negotiation teams according to their diversity. Our design decision was to consider those teams whose dissimilarity was greater than, or equal to $\bar{dt}+1.5\sigma$ as very dissimilar, and those teams whose dissimilarity was lower than, or equal to $\bar{dt}-1.5\sigma$ as very similar. In each case, 100 random negotiation teams were selected for the tests, that is, 100 teams were selected to represent the very similar team case, and 100 teams were selected to represent the very dissimilar team case. These teams participate in the different environmental scenarios, where they are confronted with one random half of all of the possible individual opponents. Therefore, each environmental scenario (complete instantiation of all the environmental conditions) consists of 100$\times$12$\times$4=4800 different negotiations (each negotiation is repeated 4 times to capture stochastic variations in the different intra-team strategies).

\subsubsection{Environment Condition: Opponent Concession Speed}
The concession speed of the opponent during the negotiation process $\beta_{op}$ may determine the final quality of the agreement for team members. For instance, if the opponent concedes very quickly towards its reservation utility, better agreements for the team may come earlier in the negotiation process. In those cases, even intra-team strategies that guarantee less degree of unanimity may achieve good results. We divided the family of concession speeds based on the classic classification of time-tactics: we considered that when $\beta_{op}=U[0.1,0.49]=VB$ the concession speed is very boulware, when $\beta_{op}=U[0.5,0.99]=B$ the concession speed is boulware, when $\beta_{op}=U[1,10]=C$ the concession speed is conceder, when $\beta_{op}=U[11,40]=VC$ the concession speed is very conceder. Similarly, when we refer to $\beta_{A}$ (the team concession speed), we will also employ the same partition in boulware (B), very boulware (VB), conceder (C), and very conceder (VC).

\subsubsection{Environment Condition: Number of Team Members}
We think that the number of team members may also influence the performance of the different intra-team strategies. Some of the strategies may become too demanding when the number of team members increases and it may result in more negotiations ending in failure. Therefore, we decided to study the effect of the team size on the performance of the different intra-team strategies. The number of team members $|A|$ ranged from 4 to 8. This number of team members  is motivated by the negotiation case employed in our experiments. We consider that groups of friends from 4 to 8 persons are reasonable in practice. 

\subsubsection{Team Performance: Number of Negotiation Rounds}
The number of negotiation rounds considers the number of interactions between the team and the opponent. It is a measure employed to assess the  negotiation time employed by the different negotiation strategies to reach a final agreement. In our study, every pair offer/counter-offer in the negotiation thread is considered as a negotiation round. In equal conditions of utility performance, those intra-team strategies that spend less negotiation rounds are preferred since they employ less negotiation time to reach a final agreement.

\subsubsection{Team Performance: Minimum Utility of Team Members}
The minimum utility of team members (Min.) in a negotiation represents the utility of the final agreement for the less benefited team member. If the final agreement is $X$ and the team is composed of M different team members $A=\{a_{1}.a_{2},...,a_{M}\}$, the minimum utility of team members can be calculated as:
\begin{equation}
Min.(X)=\underset{1\leq i \leq M}{\mbox{min}}U_{a_{i}}(X)
\end{equation}
In applications where there is a strong bond among team members (i.e., the group of travelling friends), team members may attempt to maximize the minimum utility of team members in order to avoid extremely unsatisfied team members and  a degradation of the relationship among team members. Even if a strong bond is not present among team members, an agent may attempt to maximize the minimum utility of team members if it thinks that its own utility is going to be the less favored utility by the final agreement.
\subsubsection{Team Performance: Average Utility of Team Members}
If the final agreement is $X$ and the team is composed of M different team members $A=\{a_{1}.a_{2},...,a_{M}\}$, the average utility of team members can be calculated as:
\begin{equation}
Ave.(X)=\frac{1}{M}\underset{1\leq i \leq M}{\sum}U_{a_{i}}(X)
\end{equation}
A less conservative agent may attempt to maximize the average utility of team members if it thinks that its own utility is not going to be the less favored utility by the final agreement.

\subsection{Results}
\subsubsection{Number of Negotiation Rounds}

Although we measured the number of negotiation rounds in each experiment, we found that a general pattern was found in almost every experiment. Thus, instead of commenting the results for the number of negotiation rounds in each experimental section, we decided to present the performance of the four intra-team strategies according to the number of negotiation rounds just once. As a sample for this behavior, we can observe the number of negotiation rounds spent by each intra-team strategy when team and opponent have a long deadline ($T_{op}=L$ and $T_{A}=L$), the number of team $|A|$ members is set to 4, and the opponent uses different concessions speeds $\beta_{op}$ in Table \ref{rounds}.

As long as the concession speed of the four intra-team strategies can be categorized as the same type, \textit{RE is usually the fastest intra-team strategy in number of negotiation rounds, followed by SSV, then SBV, and finally FUM}. Since less unanimity is guaranteed among team members, it is logical that there may be less conflict with the opponent and, thus, agreements are found faster with low unanimity strategies like RE and SSV. The main exception for this rule is when team members are very similar and the opponent uses either boulware or very boulware concession speeds. In those cases, FUM is able to finalize negotiations successfully in fewer rounds than SBV (and sometimes SSV). The learning heuristic employed by FUM benefits from the fact that the opponent usually concedes more in those attributes that are less important and, thus, it is able to infer a proper agenda and propose better offers to the opponent (ending the negotiation faster). This pattern did not exist when team members are very dissimilar, since in that case, FUM also has to deal with more intra-team conflict. This results in more demanding offers to guarantee unanimity. 

Additionally, as expected, \textit{as the concession strategy of team members becomes more conceder, the number of negotiation rounds spent is lower}. Thus, RE using $\beta_{A}=VB$ is slower than RE using $\beta_{A}=B$, which is slower than RE using $\beta_{A}=C$, which is slower than RE using $\beta_{A}=VC$. 

\begin{table}[t]
\center
\scalebox{0.695}{
\hspace{-1.5cm}
 \begin{tabular}{| l | l | l | l | l | l | l | l | l | l |}
\hline
\multicolumn{5}{|c|}{\textbf{Very Similar, $T_{op}=T_{A}=L$, $M=4$} } &  \multicolumn{5}{|c|}{\textbf{Very Dissimilar, $T_{op}=T_{A}=L$, $M=4$} }\\
\hline
&\textbf{$\beta_{op}=VC$} & \textbf{$\beta_{op}=C$} & \textbf{$\beta_{op}=B$} & \textbf{$\beta_{op}=VB$} & &\textbf{$\beta_{op}=VC$} & \textbf{$\beta_{op}=C$} & \textbf{$\beta_{op}=B$} & \textbf{$\beta_{op}=VB$} \\
\hline
& Rounds & Rounds & Rounds & Rounds & & Rounds & Rounds & Rounds & Rounds   \\
\hline
RE $\beta_{A}=VC$ & 2.01 & 2.28 & 7.25 & 19.57 & RE $\beta_{A}=VC$ & 2.03 & 2.33 & 7.78 & 19.71  \\
\hline
SSV $\beta_{A}=VC$ & 2.02 & 2.41 & 8.35 & 22.08 & SSV $\beta_{A}=VC$ & 2.00 & 2.71 & 9.97 & 24.35 \\
\hline
SBV $\beta_{A}=VC$ & 2.01 & 2.70 & 10.48 & 24.83 & SBV $\beta_{A}=VC$ & 2.05 & 3.44 & 12.99 & 27.33 \\
\hline
FUM $\beta_{A}=VC$ & 2.11 & 2.63 & 10.31 & 24,10 & FUM $\beta_{A}=VC$ & 2.90 & 4.52 & 16.29 & 30.70 \\
\hline
RE $\beta_{A}=C$ & 2.39 & 3.77 & 11.07 & 23.47 & RE $\beta_{A}=C$ & 2.28 & 3.30& 10.98 & 22.84  \\
\hline
SSV $\beta_{A}=C$ & 2.73 & 5.17 & 13.17 & 25.33 & SSV $\beta_{A}=C$ & 2.45 & 5.17 & 14.83 & 27.43\\
\hline
SBV $\beta_{A}=C$ & 3.02 & 6.18 & 15.55 & 27.32 & SBV $\beta_{A}=C$ & 2.99 & 7.12 & 18.64 & 30.08 \\
\hline
FUM $\beta_{A}=C$ & 4.09 & 6.23 & 14.01 & 26.45 & FUM $\beta_{A}=C$ & 6.54 & 10.47 & 21.13 & 32.66 \\
\hline
RE $\beta_{A}=B$ & 9.17 & 13.63 & 22.48 & 30.73 & RE $\beta_{A}=B$ & 6.57 & 10.02 & 19.94 & 29.19 \\
\hline
SSV $\beta_{A}=B$ & 15.53 & 19.99 & 26.97 & 32.95 & SSV $\beta_{A}=B$ & 12.09 & 18.26 & 26.42 & 33.52 \\
\hline
SBV $\beta_{A}=B$ & 17.96 & 22.40 & 28.88 & 34.21 & SBV $\beta_{A}=B$ & 16.50 & 22.74 & 30.54 & 35.76 \\
\hline
FUM $\beta_{A}=B$ & 20.31 & 23.25 & 25.59& 33.09 & FUM $\beta_{A}=B$ & 25.93 & 28.53 & 30.97 & 36.96 \\
\hline
RE $\beta_{A}=VB$ & 22.50 & 25.47 & 31.59 & 35.51 & RE $\beta_{A}=VB$ & 17.22  & 21.14 & 28.94 & 34.50 \\
\hline
SSV $\beta_{A}=VB$ & 28.62 & 31.44 & 35.27 & 37.22 & SSV $\beta_{A}=VB$ & 25.44 & 30.04 & 34.59 & 37.64 \\
\hline
SBV $\beta_{A}=VB$ & 31.50 & 33.21 & 36.24 & 37.80 & SBV $\beta_{A}=VB$ & 30.10 & 33.29 & 36.77 & 38.74 \\
\hline
FUM $\beta_{A}=VB$ & 32.77 & 33.67 & 33.97 & 37.15 & FUM $\beta_{A}=VB$ & 35.00 & 36.59 & 36.39 & 39.01 \\
\hline

 \end{tabular}}

\caption{The table depicts the comparison of the intra-team strategies when both parties have a long deadline. Results show the average number of rounds spent in the negotiation.}
\label{rounds}
\end{table}

The number of negotiation rounds spent by each intra-team strategy is especially interesting to select intra-team strategies when they perform equally in utility terms (minimum or average utility). For instance, if SBV and FUM tie in utility terms, a team is suggested to select SBV most of the times due to the fact that it usually requires less negotiation rounds, if SSV and SBV tie in utility terms, the team should select SSV since it usually requires less rounds than SBV, and so forth. 
 
\subsubsection{Same Type of Deadlines}
\label{parexp1}
The next set of experiments that we conducted consisted in assessing which intra-team strategies work better when both parties have the same type of deadline. More specifically, we chose those scenarios where both parties have short deadlines or long deadlines. Additionally, for each type of deadline,  we simulated scenarios where team members were either very dissimilar, or very similar, and gathered information about the minimum and average utility of team members regarding each possible strategy configuration (team concession speeds, intra-team strategies, opponent concession speeds, etc.).  The number of team members remained static at $|A|=4$.

\begin{table}[t]
\center
\scalebox{0.70}{
\hspace{-0.8cm}
 \begin{tabular}{|l | l | l | l | l | l | l | l | l | l | l | l | l|}
\hline
\multicolumn{13}{|c|}{\textbf{Very Similar, $T_{op}=T_{A}=S$, $M=4$} }\\
\hline
&\multicolumn{3}{|c|}{\textbf{$\beta_{op}=VC$}} & \multicolumn{3}{|c|}{\textbf{$\beta_{op}=C$}} & \multicolumn{3}{|c|}{\textbf{$\beta_{op}=B$}} & \multicolumn{3}{|c|}{\textbf{$\beta_{op}=VB$}} \\
\hline
& Min. & Ave. & Ro. & Min. & Ave. & Ro. & Min.& Ave. & Ro. & Min. & Ave. & Ro. \\
\hline
RE $\beta=B$ & 0.687 & \textbf{0.794} & 2.94 & 0.590 & 0.711 & 3.67 & 0.415 & 0.551 & 5.08 & 0.292 & 0.408 & 6.31 \\ 
\hline
SSV $\beta=B$ & 0.709 & \textbf{0.797} & 3.33 & 0.610 & 0.706 & 4.35 & 0.459 & 0.561 & 5.66 & 0.335 & 0.423 & 6.57 \\
\hline
SBV $\beta=B$ & \textbf{0.719} & \textbf{0.796} & 3.67 & \textbf{0.620} & 0.703 & 4.73 & \textbf{0.477} & 0.568 & 5.93 & \textbf{0.348} & 0.427 & 6.79 \\
\hline
FUM $\beta=B$ & 0.691 & 0.772 & 4.26 & \textbf{0.622} & \textbf{0.721} & 4.88 & \textbf{0.483} & \textbf{0.608} & 5.93 & \textbf{0.356} & \textbf{0.475} & 6.78 \\
\hline
\multicolumn{13}{|c|}{\textbf{Very Similar, $T_{op}=T_{A}=L$, $M=4$} }\\
\hline
&\multicolumn{3}{|c|}{\textbf{$\beta_{op}=VC$}} & \multicolumn{3}{|c|}{\textbf{$\beta_{op}=C$}} & \multicolumn{3}{|c|}{\textbf{$\beta_{op}=B$}} & \multicolumn{3}{|c|}{\textbf{$\beta_{op}=VB$}} \\
\hline
& Min. & Ave. & Ro. & Min. & Ave. & Ro. & Min.& Ave. & Ro. & Min. & Ave. & Ro. \\
\hline
RE $\beta=B$ & 0.758 & \textbf{0.853} & 9.18  & 0.667  & \textbf{0.779} & 13.637  & 0.476  & 0.617  & 22.486  & 0.327  & 0.457 & 30.735  \\ 
\hline
SSV $\beta=B$ & 0.757  & 0.833  & 15.54  & 0.671 & 0.765 & 19.995 & 0.514  & 0.629  & 26.981 & 0.372 & 0.484 & 32.960  \\
\hline
SBV $\beta=B$ & \textbf{0.774}  & 0.833  & 17.97  & \textbf{0.698}  & 0.765  & 22.409 & 0.541  & 0.628 & 28.891 & 0.402 & 0.489 & 34.220  \\
\hline
FUM $\beta=B$ & 0.752  & 0.816 & 20.32  & \textbf{0.695} & \textbf{0.770} & 23.264 & \textbf{0.606}  & \textbf{0.725}  & 25.603  & \textbf{0.442} & \textbf{0.560}  & 33.100  \\
\hline

\multicolumn{13}{|c|}{\textbf{Very Dissimilar, $T_{op}=T_{A}=S$, $M=4$} }\\
\hline
&\multicolumn{3}{|c|}{\textbf{$\beta_{op}=VC$}} & \multicolumn{3}{|c|}{\textbf{$\beta_{op}=C$}} & \multicolumn{3}{|c|}{\textbf{$\beta_{op}=B$}} & \multicolumn{3}{|c|}{\textbf{$\beta_{op}=VB$}} \\
\hline
& Min. & Ave. & Ro. & Min. & Ave. & Ro. & Min.& Ave. & Ro. & Min. & Ave. & Ro. \\
\hline
RE $\beta=B$ & 0.264 & 0.621 & 2.551  & 0.198  & 0.549 & 3.169  & 0.115  & 0.421 &  4.67 & 0.072  & 0.303 & 6.171  \\ 
\hline
SSV $\beta=B$ & 0.503  & \textbf{0.730}  & 2.983   & 0.417 & \textbf{0.662} & 4.054 & 0.295  & 0.526  & 5.74 & 0.182 & 0.371 & 6.771  \\
\hline
SBV $\beta=B$ & \textbf{0.550}  & \textbf{0.735}  & 3.659  & \textbf{0.467}  & 0.653  & 4.900 & \textbf{0.341}  & 0.526 & 6.229 & 0.213 & 0.368 & 7.050  \\
\hline
FUM $\beta=B$ & \textbf{0.554} & 0.709  & 5.103  & \textbf{0.468} & \textbf{0.657} & 5.750 & \textbf{0.348}  & \textbf{0.584}  & 6.564  & \textbf{0.236} & \textbf{0.445}  &  7.224 \\
\hline

\multicolumn{13}{|c|}{\textbf{Very Dissimilar, $T_{op}=T_{A}=L$, $M=4$} }\\
\hline
&\multicolumn{3}{|c|}{\textbf{$\beta_{op}=VC$}} & \multicolumn{3}{|c|}{\textbf{$\beta_{op}=C$}} & \multicolumn{3}{|c|}{\textbf{$\beta_{op}=B$}} & \multicolumn{3}{|c|}{\textbf{$\beta_{op}=VB$}} \\
\hline
& Min. & Ave. & Ro. & Min. & Ave. & Ro. & Min.& Ave. & Ro. & Min. & Ave. & Ro. \\
\hline
RE $\beta=B$ & 0.383 & 0.725 & 6.583 & 0.270  & 0.630 & 10.030  & 0.118  & 0.454 & 19.945 & 0.064  & 0.319 & 29.198  \\ 
\hline
SSV $\beta=B$ & 0.571  & 0.788  & 12.103   & 0.451 & \textbf{0.707} & 18.625 & 0.283  & 0.573  & 26.430 & 0.180 & 0.430 & 33.525  \\
\hline
SBV $\beta=B$ & \textbf{0.650}  & \textbf{0.797}  & 16.508  & 0.551  & \textbf{0.713}  & 22.746 & 0.373  & 0.562 & 30.552 & 0.242 & 0.416 & 35.766  \\
\hline
FUM $\beta=B$ & 0.627 & 0.756  & 25.938  & \textbf{0.564} & \textbf{0.714} & 28.536 & \textbf{0.489} & \textbf{0.724} & 30.976 & \textbf{0.318} &  \textbf{0.543}  & 36.968  \\
\hline

 \end{tabular}

 }
\caption{The table shows the comparison of the intra-team strategies when both parties have the same type of deadline. Results show the average for the minimum utility of team members (Min.), the average utility of team members (Ave.), and the number of rounds (Ro.). The results in bold font indicate those configurations that are statistically better and different (t-test $\alpha=0.05$) to the rest of configurations.}
\label{exp1}
\end{table}

The results for this experiment can be found in Table \ref{exp1}. It shows the average minimum utility of team members (Min.), the average of the average utility of team members (Ave.), and the average number of rounds (Ro.). It only shows the results for intra-team strategies using a Boulware concession speed since we found that this concession speed worked better than the rest of concession speeds. 

When both parties have a short deadline (first and third sub-table in Table \ref{exp1}), independently of team similarity, SBV $\beta=B$ and FUM $\beta=B$ are usually the best options for the minimum utility. The unanimity and semi-unanimity rules employed by this strategy make possible for the worst affected team member to ensure that its situation is better than with other strategies. As for the average utility of team members, FUM $\beta=B$ usually is the best option. The only exception for this pattern is when the opponent uses conceder strategies ($\beta_{op}=VC$ or $\beta_{op}=C$). In that case, all of the strategies perform similarly, especially when team members are very similar. For instance, we can observe that RE, SSV, SBV $\beta=B$ are the best option for the average utility of team members when the deadline is short, team members are very similar, and the opponent uses a very conceder strategy. In the same setting, but with the opponent using a conceder strategy, FUM is statistically better but the differences are not very important (less than a 1.8\%).

However,\textit{ when both parties have a long deadline to negotiate (subtables 2 and 4 in Table \ref{exp1}), FUM $\beta=B$ becomes the best choice for the minimum and average utility of team members in almost every scenario}. The only exceptions for this superiority are, again, scenarios where the opponent employs conceder strategies. For instance, when the deadline is long, team members are very dissimilar, and the opponent uses a very conceder strategy, SBV $\beta=B$ is the best intra-team strategy for the minimum and average utility of team members.

We can also observe that \textit{RE and SSV are specially affected by very dissimilar preferences' scenarios}. When team members are very similar, both strategies are capable of being close to SBV and FUM in the minimum and average utility of team members as long as the opponent plays conceder strategies. However, both intra-team strategies' results get further from those of SBV and FUM when team members are very dissimilar. These intra-team strategies are not able to tackle situations where team members have very dissimilar preferences due to the type of decision rule applied, and their use in such situations is discouraged. 

The reason why several strategies perform similarly in utility terms when the opponent plays conceder strategies is simple: Since the opponent concedes very fast in the first rounds of the negotiation process, as long as the team does not concede very fast (i.e., boulware strategy), all of the strategies are capable of finding a reasonable good agreement in the first rounds by letting the opponent concede and then accepting the opponent's offer. However, there is an additional reading that explains why strategies like FUM, which guarantees unanimity regarding team decisions, does not perform so well when the opponent uses conceder strategies. FUM relies on the assumption that the opponent concedes very little in those attributes that are important for its interests at the first rounds. However, when the concession strategy carried out by the opponent is conceder or very conceder (a more acute effect) big concessions are usually carried out at the first rounds. Thus, FUM is not able to infer an appropriate agenda. In \cite{sanchez-anguix12}, it was shown that as the agenda gets further from the real ranking of opponent preferences, the more demanding becomes the strategy. This may have a negative effect in the negotiation, since more negotiations may end in failure due to the high demands of the team. In fact a slight effect is observed in the results: when the opponent uses a boulware strategy, the percentage of successful negotiations is $94.6\%$ which is greater than the $92.6\%$ obtained when the opponent uses a conceder strategy and the $93.1\%$ obtained when the opponent uses a very conceder strategy.  

 Another issue found in the results is the difference between FUM and other strategies when the deadline is long.  FUM tends to obtain better results when the deadline is long for both parties. The differences with the other intra-team strategies become greater when compared with the short deadline scenario. The reason for this phenomenon is similar to the reason mentioned in the paragraph above. FUM is a strategy that relies on the information gathered in the negotiation process. Thus, when interactions are lesser, like when deadlines are short, the agenda inferred by the trusted mediator is less close to the ideal agenda. When the agenda deviates from the ideal agenda, offers proposed by the team are more demanding and less probable to be accepted by the opponent. As a matter of fact, the reader can notice that the difference on average between FUM $\beta=B$ in long deadline scenarios (aggregating all of the scenarios where the deadline is long) and the results obtained by FUM $\beta=B$ in short deadline (aggregating the scenarios where the deadline is short) scenarios counterpart is $7.9\%$, whereas it is $5.3\%$ for SBV, $5.4\%$ for SSV and $5.8\%$ for RE. Logically, every intra-team strategy benefits from having a longer deadline, but the results suggest that FUM benefits more than the rest of intra-team strategies due to its learning heuristic, which is based on the amount of information. 

\subsubsection{Different Types of Deadlines}

The next experiment consisted in studying the behavior of the different intra-team strategies when both parties have different types of deadline. Thus, in this case, one of the two parties has a deadline which is lower than the deadline of the other party. Clearly, the party with a lower deadline is at disadvantage with respect to the other party since it has fewer offers to send before ending the negotiation, and the pressure to accept the opponent's offers arises earlier. 

\paragraph{Short Team Deadline and Long Opponent Deadline}
First, we start by analyzing the case where the deadline of the team is shorter (short deadline) than the deadline of the opponent party (long deadline). Hence, $T_{A}=U[5,10]$ and $T_{op}=U[30,60]$. The results of this experiment can be found in Table \ref{exp2}.

In this case, the team has a shorter deadline and, thus, it should be at disadvantage with respect to the opponent. However, we can observe that when the opponent uses a conceder or very conceder strategy, the results are similar to the analogous case where both parties had a short deadline. These results can be explained due to the fact that since the opponent concedes very quickly, a good deal can be found for the team in the first rounds of the negotiation process and the team is not affected by the fact that its deadline is shorter. Nevertheless, as the opponent moves towards Boulware strategies, there is a clear negative effect on the minimum and average utility of team members: \textit{all of the strategies are affected by the fact that the team has a shorter deadline}. In the scenario where both parties have a short deadline (see Table \ref{exp1}), the average for the average utility of team members in conceder settings (aggregating those negotiations where $\beta_{op}=C$ or $\beta_{op}=VC$) is 0.67, and the average for the average utility of team members in boulware settings (aggregating those negotiations where $\beta_{op}=B$ or $\beta_{op}=VB$) is 0.45. Thus, the average utility for team members is reduced a 22\%. In this experiment (see Table \ref{exp2}), the average of the average utility of team members in conceder settings is 0.63, whereas the average of the average utility of team members in boulware settings is 0.10. Therefore, the average utility of team members is reduced a 53\%, approximately doubling the difference found in the case where both parties had a short deadline.

When \textit{team members are very similar} (upper sub-table in Table \ref{exp2}), it can be observed that, as in the scenario where both parties have a short deadline and team members are very similar,  several strategies perform very similarly. The main difference resides in the fact that the only strategy capable of reaching similar results to FUM $\beta=B$ in the minimum and average utility is RE $\beta=B$. Differently to the case when team members are very similar and the deadline for both parties is short, the RE $\beta_{A}=B$ strategy is capable of achieving similar results to the other intra-team strategies even in less conceding settings ($\beta_{op}=C$, $\beta_{op}=B$, and $\beta_{op}=VB$). These results suggest that, \textit{despite not assuring any minimum level of unanimity, employing a representative with a reasonably slow concession (boulware) leads to good results compared with those obtained by other intra-team strategies}. A closer look at the experiments threw some light over these results. For instance, when $\beta_{op}=B$, the number of successful negotiations was 2695 for RE $\beta_{A}=B$, 1925 for FUM $\beta_{A}=B$, 1855 for UBS $\beta_{A}=B$, and 2394 for SSV $\beta_{A}=B$. The average utility for successful negotiations was 0.32 for RE $\beta_{A}=B$, 0.34 for SSV $\beta_{A}=B$, 0.39 for SBV $\beta_{A}=B$, and 0.42 for FUM $\beta_{A}=B$. Hence, despite obtaining less quality results in successful negotiations, the representative approach becomes a good option for these scenarios because it leads to a great number of negotiations ending in success where other intra-team strategies fail to succeed (utility=0). SSV, UBS, and FUM need more interactions to find a satisfactory deal, but when they find it, it is better in utility terms. However, in average, a representative approach may be more adequate for settings where the team has a shorter deadline than the opponent.

As for the scenario where \textit{team members are very dissimilar} (lower sub-table in Table \ref{exp2}), we can observe that the negative effect produced by having a shorter deadline is especially acute when the opponent uses boulware or very boulware concessions. The dissimilarities between team members, and the fact that there are very few interactions to find a deal that satisfies both team and opponent, contribute to a strong reduction in the minimum and the average utility of team members. In terms of the minimum utility of team members, $FUM$ and $SBV$ $\beta_{A}=B$ work better when the opponent uses conceder or very conceder concessions. However, \textit{almost every intra-team strategy performs equally bad in terms of the minimum utility of team members when the opponent moves towards boulware concessions} (especially in the very boulware case). In this case, the representative approach can no longer compete with the rest of strategies in terms of utility in most scenarios. Nevertheless, despite team members being very dissimilar and RE not guaranteeing any unanimity regarding team decisions, RE performs slightly better than the rest in terms of the average utility of team members when the opponents concedes using boulware. The explanation to this phenomenon is similar to the case where team members were very similar: a lesser number of negotiations end in failure (26\% failures for RE, 33\% for SSV, 48\% for SBV, and 46\% for FUM), which compensates for the dissimilarity between team members' preferences and the unanimity level guaranteed by RE. In any case, the utility obtained for team members is so low in the average and minimum utility of team members that, in some cases, \textit{it may even be better not to negotiate with such kind of opponent and spend computational resources in looking for another alternative}.

\begin{table}[t]
\center
\scalebox{0.70}{
\hspace{-0.8cm}
 \begin{tabular}{|l | l | l | l | l | l | l | l | l | l | l | l | l|}
\hline
\multicolumn{13}{|c|}{\textbf{Very Similar, $T_{op}=L$, $T_{A}=S$, $M=4$} }\\
\hline
&\multicolumn{3}{|c|}{\textbf{$\beta_{op}=VC$}} & \multicolumn{3}{|c|}{\textbf{$\beta_{op}=C$}} & \multicolumn{3}{|c|}{\textbf{$\beta_{op}=B$}} & \multicolumn{3}{|c|}{\textbf{$\beta_{op}=VB$}} \\
\hline
& Min. & Ave. & Ro. & Min. & Ave. & Ro. & Min.& Ave. & Ro. & Min. & Ave. & Ro. \\
\hline
RE $\beta=B$ & 0.643 & \textbf{0.756} & 3.232  & 0.483 & \textbf{0.609} & 4.542 & \textbf{0.152} & \textbf{0.242} & 7.112 & \textbf{0.027} & \textbf{0.048} & 8.284 \\ 
\hline
SSV $\beta=B$ & 0.648 & 0.748 & 3.895 & 0.465 & 0.576 & 5.468 & \textbf{0.145} & 0.227 & 7.544 & 0.024 & 0.040 & 8.396 \\
\hline
SBV $\beta=B$ & \textbf{0.656} & 0.743 & 4.370 & 0.473 & 0.568 & 5.995 & 0.139 & 0.199 & 7.876 & 0.019 & 0.028 & 8.457 \\
\hline
FUM $\beta=B$ & \textbf{0.651} & 0.747 & 4.733 & \textbf{0.494} & \textbf{0.612} & 5.931 & \textbf{0.150} & 0.222 & 7.818 & \textbf{0.029} & \textbf{0.046} & 8.424  \\
\hline
\multicolumn{13}{|c|}{\textbf{Very Dissimilar, $T_{op}=L$, $T_{A}=S$, $M=4$} }\\
\hline
&\multicolumn{3}{|c|}{\textbf{$\beta_{op}=VC$}} & \multicolumn{3}{|c|}{\textbf{$\beta_{op}=C$}} & \multicolumn{3}{|c|}{\textbf{$\beta_{op}=B$}} & \multicolumn{3}{|c|}{\textbf{$\beta_{op}=VB$}} \\
\hline
& Min. & Ave. & Ro. & Min. & Ave. & Ro. & Min.& Ave. & Ro. & Min. & Ave. & Ro. \\
\hline
RE $\beta=B$ & 0.245 & 0.596 & 2.771  & 0.153 & 0.482 & 3.870 & 0.025 & \textbf{0.170} & 7.111 & \textbf{0.002} & 0.040 & 8.231 \\ 
\hline
SSV $\beta=B$ & 0.459 & 0.703 & 3.444 & 0.280 & 0.540 & 5.553 & \textbf{0.035} & 0.156 & 7.919 & \textbf{0.002} & 0.017 & 8.526 \\
\hline
SBV $\beta=B$ & \textbf{0.511} & \textbf{0.706} & 4.377 & 0.313 & 0.496 & 6.677 & 0.028 & 0.082 & 8.282 & 0.001 & \textbf{0.064} & 8.560 \\
\hline
FUM $\beta=B$ & \textbf{0.520} & \textbf{0.704} & 5.825 & \textbf{0.336} & 0.545 & 7.003 & 0.026 & 0.084 & 8.333 & 0.001 & \textbf{0.060} & 8.565  \\
\hline
\end{tabular}

 }
\caption{Comparison of the intra-team strategies when the team has a short deadline and the opponent party has a long deadline. Results show the average for the minimum utility of team members (Min.), the average utility of team members (Ave.), and the number of rounds (Ro.). The results in bold font indicate those configurations that are statistically better and different (t-test $\alpha=0.05$) to the rest of configurations.}
\label{exp2}
\end{table}

\begin{figure}[!h]
 \begin{center}
\scalebox{1.2}{
\hspace{-1.2cm}
 \includegraphics[width=\linewidth]{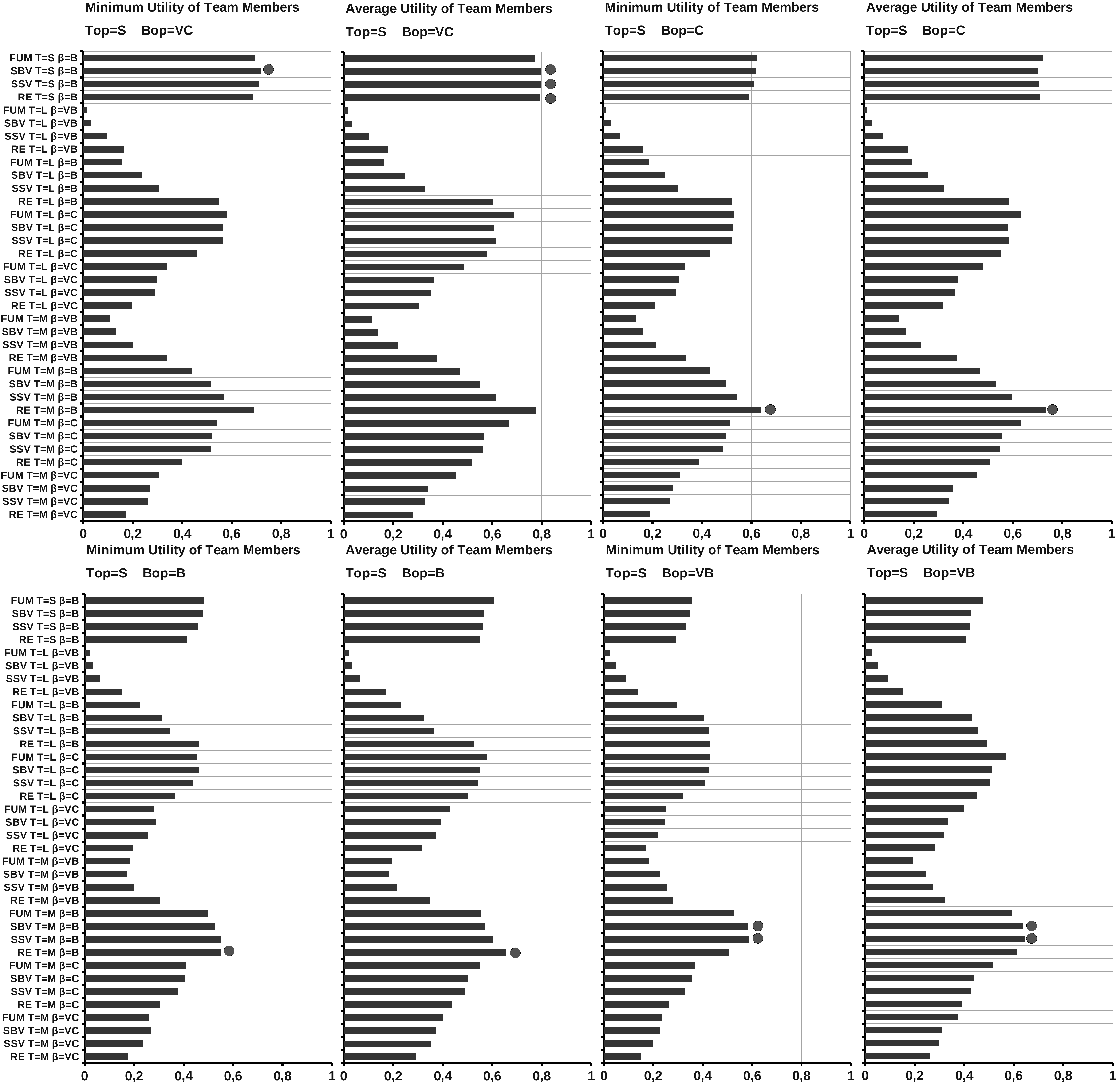}}
\end{center}
\caption{Results for very similar team members when the team has a long deadline and the opponent has a short deadline. The dots indicate those configurations that perform statistically better than the rest (t-test $\alpha=0.05$)}
\label{exp:more:similar}
\end{figure}

\begin{figure}[!h]
 \begin{center}
\scalebox{1.2}{
 \hspace{-1.2cm}
 \includegraphics[width=\linewidth]{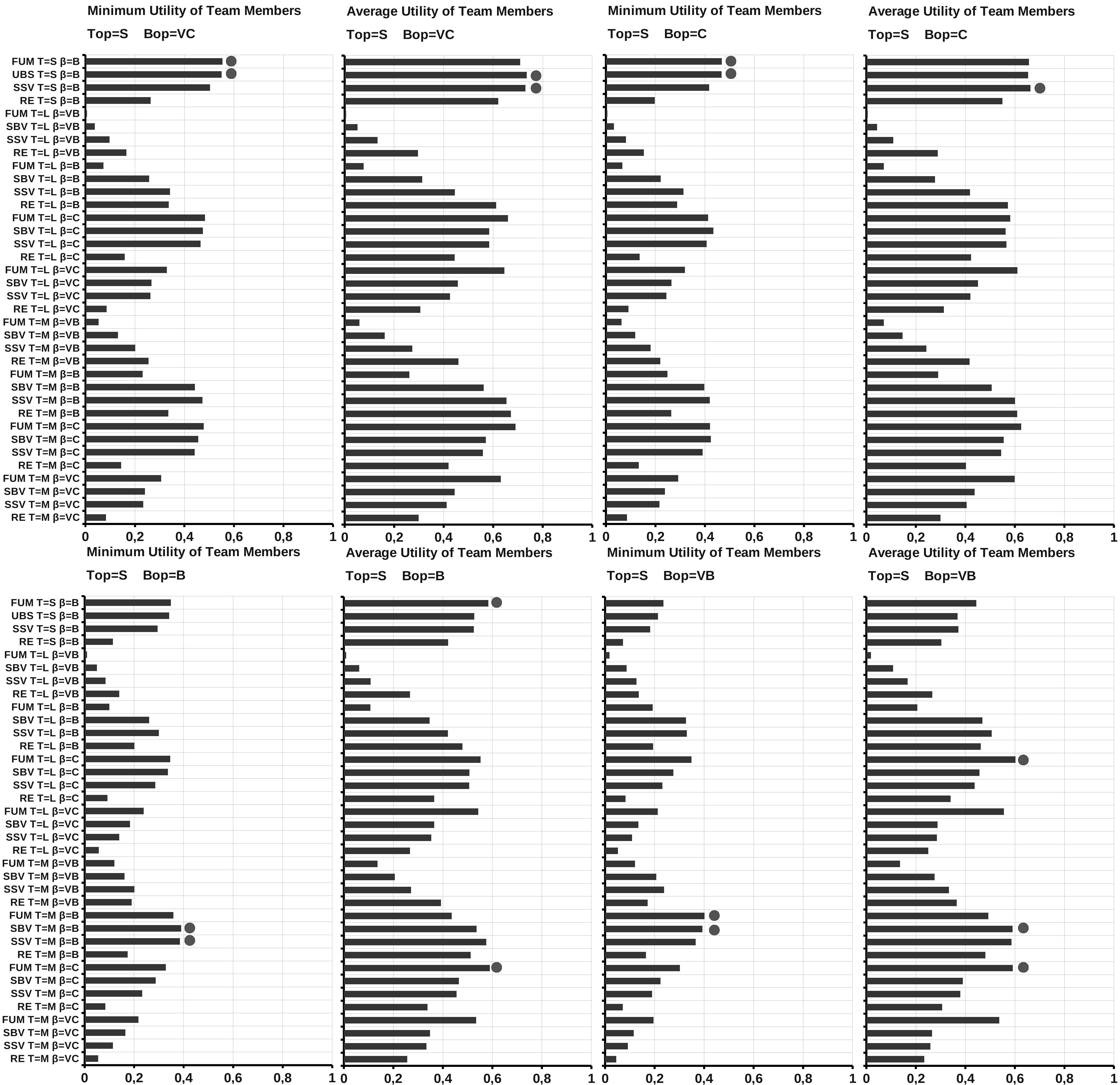}}
\end{center}
\caption{Results for very dissimilar team members when the team has a long deadline and the opponent has a short deadline. The dots indicate those configurations that perform statistically better than the rest (t-test $\alpha=0.05$)}
\label{exp:more:dissimilar}
\end{figure}

\paragraph{Long Team Deadline and Short Opponent Deadline}
In this case, the team has an advantage over the opponent since its maximum deadline is longer than the opponent's deadline. The goal of these experiments is to determine the combination of intra-team strategies and negotiation parameters that maximize the different social welfare measures employed. Thus, if the team has a maximum deadline equal to the uniform distribution $T_{A}=U[30,60]$, the team may decide to play (prior to the negotiation) a different class of deadline like a medium deadline ($T_{A}=U[11,29]$) or a short deadline ($T_{A}=U[5,10]$) if the results of the simulation suggest that better results are obtained by not playing the maximum deadline. Thus, we also show the results for teams that play a medium deadline, and teams that play a short deadline. In this experiment, the opponent plays a short deadline $T_{op}=U[5,10]$. The results of this experiment for the very similar scenario can be observed in Fig. \ref{exp:more:similar}, whereas the results for the very dissimilar scenario can be observed in Fig. \ref{exp:more:dissimilar}.

We start by analyzing the results for scenarios where team members are very similar (Fig. \ref{exp:more:similar}). We can observe that for situations where the opponent is very conceder, the team benefits from playing strategies with the same deadline. Since the opponent concedes very fast in the first negotiation rounds, the best deals for the team may be proposed in the first negotiation rounds. Playing a longer deadline may be risky since the team may have extremely high aspirations during the whole negotiation, which results in most offers being rejected and ending the negotiation in failure. As a matter of fact, the number of successful negotiations for intra-team strategies playing a short deadline and boulware concession was 95.1\%, 68\% for medium deadline and boulware concession, and 45\% for long deadline and boulware concession, 29\% for medium deadline and very boulware concession, and 14\% for long deadline and very boulware concession. Other configurations may have a higher number of successful negotiations, but they are not able to retain as much utility as the boulware configuration. As the opponent starts to move towards strategies that concede more slowly, the best intra-team strategies for the team are those played with a medium deadline and boulware strategy (RE, SBV and SSV $\beta=B$). In those cases, the opponent may not propose the best deals for the team until its last negotiation rounds. Thus, playing a slightly longer deadline with a boulware concession comes at an advantage for the team since the team does not fully concede in the whole negotiation and still accepts last opponent's offers. Some strategies played with a medium deadline like FUM $\beta=B$ are still too demanding, end up in more negotiation failures, and have very little information to learn the opponents' preferences.  

The very dissimilar scenario (Fig. \ref{exp:more:dissimilar}) is a little bit different. In this scenario, the team needs to deal with strong divergences in their preferences too. Thus, teams are prone to be more demanding in order to accommodate the preferences of as many team members as possible. We can observe that for cases where the opponent uses conceder strategies, the team should play boulware strategies with the same deadline. Similarly to the very similar scenario, playing a longer deadline is risky since it results in extremely high aspirations and most offers being rejected. However, in the very dissimilar scenario, the transition from selecting short deadline strategies to selecting medium deadline strategies does not appear until the opponent uses boulware strategies. This may be explained precisely due to the dissimilarity among team members, which requires stronger demands that are not met when playing medium deadline. As the opponent starts to concede using boulware strategies, the best intra-team strategies are usually found in the medium deadline, as in the very similar scenario case.

In conclusion, in this experiment we have observed that, generally, \textit{even though the team is able to play a long deadline and the opponent plays a short deadline, the team would benefit more from playing the same type of deadline than the opponent or a slightly longer deadline}.

\subsubsection{Team size effect on intra-team strategies}

\begin{figure}[!h]
 \begin{center}
 \scalebox{1.45}{
\hspace{-2.1cm}
 \includegraphics[width=\linewidth]{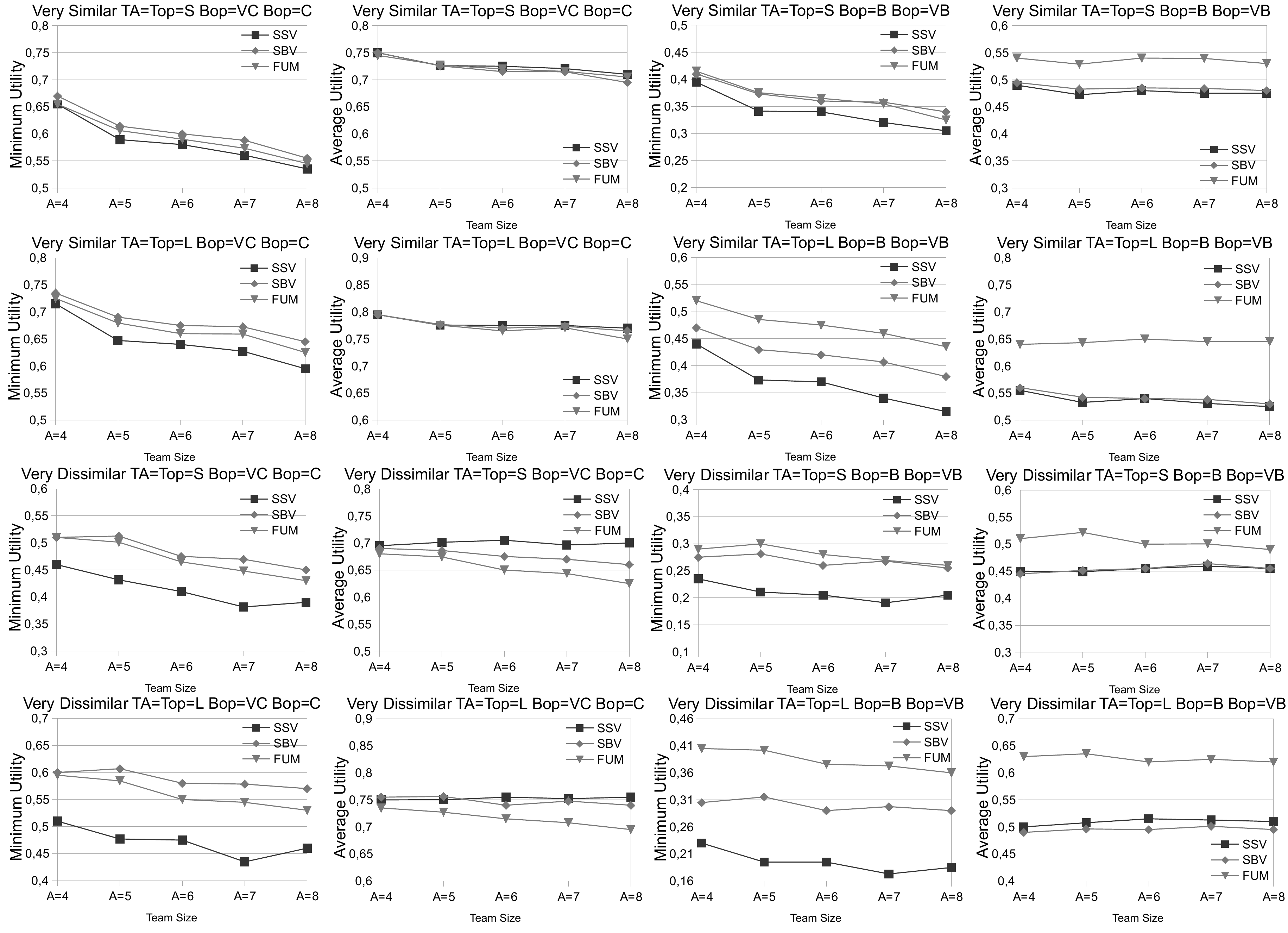}}
\end{center}
\caption{Size effect when team and opponent have the same type of deadline}
\label{exp:teamsize}
\end{figure}

We also decided to analyze the effect of the team size on the performance of the different intra-team strategies. Thus, we repeated the conditions in \ref{parexp1}  increasing the number of team members. However, we only analyzed intra-team strategies whose $\beta_{A}=B$ since they were those one that obtained better results in Table \ref{exp1}. We excluded the RE strategy from the analysis. Since team members do not interact in RE and no unanimity level is guaranteed, the inclusion of additional team members should not affect the way in which the strategy works.  The results of this experiment can be found in Figure \ref{exp:teamsize}. It shows the average and minimum utility of team members for teams of size $|A|=\{4,5,6,7,8\}$.

Generally, it can be observed in all of the graphics in Figure \ref{exp:teamsize} that, as the number of team members increases, \textit{the quality of the results in terms of the minimum and the average utility is reduced}.  This behavior was expected since as the number of agents increases, the set of possible agreements is reduced and the conflict inside the team and with the opponent is increased. However, the reduction in utility terms can be appreciated more easily in the minimum utility of team members. The average for the average utility of team members when $|A|=4$ is 0.70 (aggregating all other factors) and 0.67 for $|A|=8$ (aggregating all other factors), whereas the average for the minimum utility of team members when $|A|=4$ is 0.48 and 0.41 for $|A|=8$. As the number of team members increases, the contribution of each team member to the average utility is lesser, and that is the reason why the negative effect of team size on utility measures can be observed more easily in the minimum utility of team members than in the average utility of team members. 

We expected that as the number of team members increased, the performance of unanimity intra-team strategies like FUM would greatly decrease compared to the performance of SSV since more team members would increase the demands of the team and make offers less interesting for the opponent. However, the difference in performance between the three strategies is approximately maintained in almost every graphic as the number of team members increases. Therefore, \textit{team size did not have a different effect on the performance of the three intra-team strategies, affecting all of intra-team strategies equally}. The decision on which intra-team strategy should be chosen is not affected by team size.

The only clear exceptions to this rule are scenarios where the opponent uses conceder strategies ($\beta_{op}=C$ and $\beta_{op}=VC$) and team members' preferences are very dissimilar (first two graphics in rows 3 and 4, Figure \ref{exp:teamsize}). In these scenarios, we can observe that there is a special negative effect of team size on the performance (mininum utility and specially in the average utility) of FUM with respect to the other intra-team strategies, which results in FUM being one of the worst choices when the number of team members in large (e.g., second graphics in rows 3 and 4, Figure \ref{exp:teamsize}). As a numeric example of the reduction in the performance of FUM , the difference in the average utility between SBV and FUM goes from approximately a $2\%$ ($|A|=4$) to $10\%$ ($|A|=8$) when $\beta_{op}=VC$ and the deadline is short, from approximately a $0\%$ ($|A|=4$) to $5\%$ ($|A|=8$) when the deadline is short and $\beta_{op}=C$, and from $3\%$ ($|A|=4$) to $8\%$ ($|A|=8$) when the deadline is long and $\beta_{op}=VC$ . This phenomenon has a reasonable explanation. When the opponent uses conceder strategies, FUM has greater difficulties to learn a proper attribute agenda. If the number of team members increases and they are very dissimilar, the demands of team members increase, which summed up to the fact that the agenda does not properly reflect the preferences of the opponent, results in demanding team proposals.

\section{Related Work}
\label{sec:related}
 Multi-agent systems have gained a growing interest as the infrastructure necessary for the next generation of distributed systems. Due to the inherent conflict among agents, techniques that allow agents to solve their conflicts and cooperate are needed. This need is what has given birth to a group of technologies which have recently been referred to as agreement technologies \cite{luck08,sierra11}. Trust and reputation \cite{sabater05,such11,such12}, norms \cite{dignum99,criado11}, agent organizations \cite{horling04,esparcia11,silva12}, argumentation \cite{rahwan03,pajares11} and automated negotiation \cite{jennings01,sanchez-anguix11b} are part of the core that makes up this new family of technologies.

Even though agreement technologies are a novel topic in the community of agent research, some of its core technologies like automated negotiation have been studied by scholars for a few years. In definition, automated negotiation is a process carried out between two or more parties in order to reach an agreement by means of exchange of proposals. Two different research trends can be distinguished in automated negotiation models. The first type of model aims to calculate the optimum strategy given certain information about the opponent and the negotiation environment \cite{serrano03,digiunta06,fatima06}. The second type of model encloses heuristics that do not calculate the optimum strategy but obtain results that aim to be as close to the optimum as possible \cite{faratin98,jonker01,faratin02,lai08}. These models assume imperfect knowledge about the opponent and the environment, and aim to be computationally tractable while obtaining good results. This present work can be classified into the latter type of models.

In multi-agent systems, most of the research has concentrated on bilateral models where each party is a single individual. The present article studies bilateral negotiations where at least one of the parties is a negotiation team, composed by more than a single individual. It should be noted that the problem of finding an agreement for a negotiation team is inherently complex since it not only requires finding an agreement with the other party but it also entails reaching some type of unanimity within the team. Even though communications with the opponent party may be similar to classical bilateral models, negotiation teams may require an additional level of negotiation that involves team members. Thus, classical bilateral models cannot be applied directly if a certain level of unanimity regarding team decisions is necessary . As far as we know, our previous work \cite{sanchez-anguix10,sanchez-anguix11,sanchez-anguix12,sanchez-anguix12b} is the only work that focuses on negotiation teams.

 In \cite{sanchez-anguix10} we introduced the topic of negotiation teams in agent research from a descriptive perspective. We analyzed the different phases necessary for an agent-based negotiation team to face such negotiations with success. Apart from the phases that we identified, we also described the current technologies that may be appropriate for the development of such phases. Later, we introduced our first experimental study \cite{sanchez-anguix11} comparing intra-team strategies in different negotiation environments. That paper should be considered the preliminary basis for our current analysis. We have introduced changes in the intra-team strategies, and our current study applies a more fine-grained analysis of the negotiation environment and its possible scenarios. Additionally, we also studied the properties of the Full Unanimity Mediated intra-team strategy in \cite{sanchez-anguix12}. However, a thorough analysis of how environmental conditions affect team performance was not carried out. Finally, we should also highlight our work regarding the study of cultural factors in negotiation teams \cite{sanchez-anguix12b}. The setting is different to the current article. We attempted to propose a computational model for explaining how human cultural factors affect team dynamics in negotiation teams composed by humans. In this present article we do not consider humans but automated agents. Therefore, human factors are not relevant to the present study.

Apart from agent-based negotiation teams, bilateral negotiation is perhaps the most similar topic to our current research. Hence, we describe some of the most important bilateral models that assume imperfect knowledge.  Faratin et al. \cite{faratin98} propose a bilateral negotiation model for service negotiation where agents apply and mix different concession tactics (i.e., time-dependent, imitative and resource-dependent).  In their work, they analyze the impact of the model's parameters and determine which configurations work better in different scenarios by means of experiments. Our proposed work also assumes the use of time-dependent concession tactics for the calculation of agents' aspirations at each negotiation round. Additionally, we also take an experimental approach to validate the impact of our model's parameters. Later, the authors proposed a bilateral negotiation model \cite{faratin02} whose main novelty was the use of trade-offs to improve agreements between two parties. A trade-off consists of reducing the utility obtained from some negotiation issues with the goal of obtaining the same exact utility from other negotiation issues. The rationale behind trade-offs is to make the offer more likable for the opponent while maintaining the same level of satisfaction for the proposing agent. For that purpose, the authors propose a fuzzy similarity heuristic that proposes the most similar offer to the last offer received from the opponent. Some of our intra-team strategies like Similarity Simple Voting and Similarity Borda Voting also employ similarity heuristics to attempt to satisfy team members' preferences and the opponent's preferences.

Jonker and Treur propose the Agent-Based Market Place (ABMP) model \cite{jonker01} where agents, engage in bilateral negotiations. ABMP is a negotiation model where proposed bids are concessions to previous bids. The amount of concession is regulated by the concession factor (i.e., reservation utility), the negotiation speed, the acceptable utility gap (maximal difference between the target utility and the utility of an offer that is acceptable), and the impatience factor (which governs the probability of the agent leaving the negotiation process). 

Lai et al. \cite{lai08} propose a decentralized bilateral negotiation model where agents are allowed to propose up to $k$ different offers at each negotiation round. Offers are proposed from the current iso-utility curve according to a similarity mechanism that selects the most similar offer to the last offer received from the opponent. The selected similarity heuristic is the Euclidean distance since it is general and does not require domain-specific knowledge and information regarding the opponent's utility function. Results showed that the strategy is capable of reaching agreements that are very close to the Pareto Frontier. Sanchez-Anguix et al. \cite{sanchez-anguix11b}  proposed an enhancement for this strategy in environments where  computational resources are very limited and utility functions are complex. It relies on genetic algorithms to sample offers that are interesting for the agent itself and creates new offers during the negotiation process that are interesting for both parties. Results showed that the model is capable of obtaining statistically equivalent results to similar models that had the full iso-utility curve sampled, while being computationally more tractable. As commented above, some of our intra-team strategies use similarity heuristics to satisfy team members' preferences and the opponent's preferences. 

Another topic that resembles team negotiations are multi-party negotiations. Several works have been proposed in the literature along this line \cite{ehtamo01,klein03,ito10}. For instance, Ehtamo et al. \cite{ehtamo01} propose a mediated multi-party negotiation protocol which looks for joint gains in an iterated way. The algorithm starts from a tentative agreement and moves in a direction according to what the agents prefer regarding some offers' comparison. Results showed that the algorithm converges quickly to Pareto optimal points. Klein et al. \cite{klein03} propose a mediated negotiation model which can be extended to multiple parties. Their main goal is to provide solutions for negotiation processes that use complex utility functions to model agents' preferences. The negotiation attributes are not independent. Therefore, preference spaces cannot be explored as easily as in the linear case. Later, Ito et al. \cite{ito10} proposed different types of utility functions (cube and cone constraints) and multiparty negotiation models for such utility functions. The main difference between our work and multi-party negotiations lies in the nature of the conflict and how protocols are devised. Even though each team member could be viewed as a participant in a multi-party negotiation with the opponent, it is natural to think that team members' preferences are more similar (e.g., a team of buyers, a group of friends, etc.) and they trust other teammates more than the opponent (i.e., they may share more information). Furthermore, multi-party negotiation models may be unfair for agents that are alien to the team if the number of team members exceeds the number of other participants. In that case, multi-party models may be inclined to move the negotiation towards agreements that maximize the preferences of most participants (i.e., team members).

Multi-agent teamwork is also a close research area. Agent teams have been proposed for a variety of tasks such as Robocup \cite{stone99}, rescue tasks \cite{kitano01}, and transportation tasks \cite{jennings95}. However, as far as we know, there is no published work that considers teams of agents negotiating with an opponent. Most works in agent teamwork consider fully cooperative agents that work to maximize shared goals. The team negotiation setting is different since, even though team members share a common interest related to the negotiation, there may be competition among team members to maximize one's own preferences.

\section{Conclusions and Future Work}
\label{sec:conclusions}
An agent-based negotiation team is a group of two or more interdependent agents that join together as a single negotiation party because they share some common interests in the negotiation at hand. Intra-team strategies govern which decisions are taken by the negotiation team, and how and when these decisions are taken. The goal of this article is studying how environmental conditions affect the performance of different intra-team strategies for a team negotiating with an opponent. We studied how the deadline of both parties, the concession speed of the opponent, similarity among team members' preferences and team size affect the performance of Representative (RE) intra-team strategy, Similarity Simple Voting (SSV) intra-team strategy, Similarity Borda Voting (SBV) intra-team strategy and Full Unanimity Mediated (FUM) intra-team strategy in terms of the minimum utility of team members, the average utility of team members and the number of negotiation rounds. \textit{The results suggest that depending on the environmental conditions and the team performance metric, team members should select different intra-team strategies}, which confirms our initial hypothesis in this article. Next, we summarize some of the most important results found in this paper:
\begin{itemize}
 \item Generally, when the concession speed is the same for the different intra-team strategies, RE takes less numbers of negotiation rounds than SSV, which takes less number of rounds than SBV, which takes less number of rounds than FUM. The exception for this rule is when team members are very similar and the opponent uses boulware or very boulware strategies, which makes FUM usually faster than SBV.
 \item FUM tends to clearly outperform the rest of intra-team strategies studied in utility terms (minimum and average utility of team members) when the deadline of both parties is long and the opponent uses either boulware of very boulware concession strategies. When the opponent uses conceder or very conceder strategies, different intra-team strategies tie in terms of the minimum and average utility of team members depending on the rest of environmental conditions.
 \item When the team deadline is way shorter than the opponent's deadline, all of the intra-team strategies are negatively affected in the results obtained in the minimum and average utility of team members. Additionally, if team members are very similar, RE becomes one of the best choices for the average utility of team members since it is capable of ending more negotiations successfully where other intra-team strategies fail. If team members are very dissimilar, FUM and SBV tend to work better in terms of utility (minimum and average). However, if the opponent uses boulware or very boulware concession strategies every intra-team strategy performs equally bad and team members are encouraged to look for other negotiation alternatives.
 \item In situations where the team's maximum deadline is longer than the opponent's deadline, the team should not play intra-team strategies with the maximum deadline but intra-team strategies with the same type of deadline than the opponent or a slightly longer type of deadline. Otherwise, the team performance in utility terms is not maximized due to more negotiations ending in failure.
 \item As the number of team members increases, the performance in utility terms of all of the intra-team strategies is negatively affected. However, in general, all of the intra-team strategies studied are equally affected by the increment in the number of team members. Thus, team size did not have an effect on the intra-team strategy that should be selected by team members to maximize the minimum or the average utility of team members.
\end{itemize}

The field of negotiation teams is novel in the area of multi-agent systems. Therefore, there is much work to be done in order to advance the state-of-the-art. Current works in agent-based negotiation teams \cite{sanchez-anguix10,sanchez-anguix11,sanchez-anguix12,sanchez-anguix12b} have focused on negotiation processes where the team has a strong potential for cooperation since team members share the same type of monotonic valuation function for negotiation issues. However, it is possible to assume that in some negotiation scenarios there is more conflict among team members since valuation functions may be of a different type of monotonic function among team members, or the valuation function itself is not predictable in the negotiation domain (e.g., colors, brands, etc.). Our current future work involves designing intra-team strategies that are able to tackle negotiation domains where attribute's valuation functions may be unpredictable. RE, SSV and SBV are able to handle such types of domains by definition. However, FUM, which is the strategy capable of guaranteeing unanimity regarding team decisions, is not capable of handling domains where attributes are unpredictable (due to the max/min aggregation operator). Hence, our future works consists in proposing an intra-team strategy capable of guaranteeing unanimity for negotiation domains where attributes may not be predictable.

On the other hand, since the results of this present article have shown that environmental conditions do affect the performance of intra-team strategies, we plan to propose a mechanism that allows team members to infer the most probable state of the negotiation environment, and according to that information, advise the use of an appropriate intra-team strategy.

Finally, in our current work we assume that team members have the same knowledge about the negotiation domain and they have the same skills. It may be interesting to study scenarios where team members have different knowledge and skills.

\section*{Acknowledgements}
This work is supported by TIN2011-27652-C03-01, TIN2009-13839-C03-01, CSD2007-00022 of the Spanish government, and FPU grant AP2008-00600 awarded to Víctor Sánchez-Anguix. We would also like to thank anonymous reviewers and assistants of AAMAS 2011 who helped us to improve our previous work, making this present work possible.

\bibliographystyle{model1b-num-names}      
\bibliography{NT-ISFinal}   

\end{document}